\documentclass[useAMS,usenatbib]{mnras}

\usepackage{mathptmx}
\usepackage{mathtools}
\usepackage[T1]{fontenc}
\usepackage{txfonts}
\usepackage{ae,aecompl}
\usepackage{soul}
\usepackage{array}
\usepackage{amssymb}
\usepackage{pdflscape}
\usepackage{CJKutf8}

\usepackage[T1]{fontenc}
\usepackage{ae,aecompl}

\usepackage{soul}
\usepackage{indentfirst}
\usepackage{array}
\usepackage{times}

\usepackage{float}
\usepackage{graphicx}	

\usepackage{amsmath}	
\usepackage{amssymb}	
\usepackage{threeparttable}
\usepackage{subfigure}
\usepackage[all]{hypcap}

\usepackage[usenames]{color}
\usepackage{pdflscape}
\def\mnras{MNRAS}
\def\apjl{ApJL }
\def\aj{AJ }
\def\apj{ApJ }

\def\pasp{PASP }

\def\apjs{ApJS }
\def\araa{ARAA }
\def\aap{A\&A }
\def\nat{Nature }
\hypersetup{colorlinks=true,citecolor=blue,linkcolor=blue,filecolor=black,runcolor=black,breaklinks=true}
\definecolor{purple}{RGB}{160,32,240}
\definecolor{Cerulean}{RGB}{0,123,167}
\usepackage{etoolbox}

\newcommand{\mbh}{$M_{\bullet}$}
\newcommand{\mstar}{$M_{*}$}
\newcommand{\Msun}{M_{\odot}}

\newcommand{\mpeak}{$M_\mathrm{peak}$}

\newcommand{\shadedregions}{The shaded regions show the 68\% confidence intervals inferred from the model posterior distribution.}

\newcommand{\bhbm}{$M_\bullet$--$M_\mathrm{bulge}$}
\newcommand{\bhsm}{$M_\bullet$--$M_*$}

\title[\textsc{Trinity} IV: Galaxy--SMBH Connection at $z\gtrsim 7$]{\textsc{Trinity} IV: Predictions for Supermassive Black Holes at $z\gtrsim 7$}

\author[H. Zhang et al.]{
Haowen Zhang (\begin{CJK*}{UTF8}{gbsn}
张昊文
\end{CJK*}),$^{1}$\thanks{E-mail: hwzhang0595@arizona.edu}
Peter Behroozi,$^{1,2}$
Marta Volonteri,$^{3}$
Joseph Silk,$^{3,4,5}$
\newauthor{
Xiaohui Fan,$^{1}$
James Aird,$^{6,7}$
Jinyi Yang (\begin{CJK*}{UTF8}{gbsn}
杨锦怡
\end{CJK*}),$^{1,\ast}$
Feige Wang (\begin{CJK*}{UTF8}{gbsn}
王飞格
\end{CJK*}),$^{1}$}
\newauthor{
and Philip F. Hopkins$^{8}$
}
\\
$^{1}$University of Arizona, 933 N Cherry Ave., Tucson, AZ 85721, USA \\
$^{2}$Division of Science, National Astronomical Observatory of Japan, 2-21-1 Osawa, Mitaka, Tokyo 181-8588, Japan\\
$^{3}$Institut d'Astrophysique de Paris (UMR 7095: CNRS \& Sorbonne Universite), 98 bis Bd. Arago, F-75014, Paris, France\\
$^{4}$Department of Physics and Astronomy, Johns Hopkins University, Baltimore, MD 21218, USA\\
$^{5}$BIPAC, Department of Physics, University of Oxford, Keble Road, Oxford OX1 3RH, UK\\
$^{6}$Institute for Astronomy, University of Edinburgh, Royal Observatory, Edinburgh EH9 3HJ, UK\\
$^{7}$Department of Physics and Astronomy, University of Leicester, University Road, Leicester LE1 7RH, UK\\
$^{8}$Theoretical Astrophysics, California Institute of Technology, Pasadena, CA 91125, USA\\
$^{\ast}$Strittmatter Fellow
}

\date{Accepted XXX. Received YYY; in original form ZZZ}

\pubyear{2020}

\begin{document}
\label{firstpage}
\pagerange{\pageref{firstpage}--\pageref{lastpage}}
\maketitle

\begin{abstract}
We present predictions for the high-redshift halo--galaxy--supermassive black hole (SMBH) connection from the \textsc{Trinity} model. Constrained by a comprehensive compilation of galaxy ($0\leq z \leq 10$) and SMBH datasets ($0\leq z \leq 6.5$), \textsc{Trinity} finds: 1) The number of SMBHs with $M_\bullet > 10^9 M_\odot$ in the observable Universe increases by six orders of magnitude from $z\sim 10$ to $z\sim 2$, and by another factor of $\sim 3$ from $z\sim 2$ to $z=0$; 2) The $M_\bullet>10^9/10^{10} M_\odot$ SMBHs at $z\sim6$ live in haloes with $\sim(2-3)/(3-5)\times 10^{12} M_\odot$; 3) the new JWST AGNs at $7\lesssim z \lesssim 11$ are broadly consistent with the median SMBH mass--galaxy mass relation for AGNs from \textsc{Trinity}; 4) Seeds from runaway mergers in nuclear star clusters are viable progenitors for the SMBHs in GN-z11 ($z=10.6$) and CEERS\_1019 ($z=8.7$); 5) $z=6-10$ quasar luminosity functions from wide area surveys by, e.g., \textit{Roman} and \textit{Euclid}, will reduce uncertainties in the $z=6-10$ SMBH mass--galaxy mass relation by up to $\sim 0.5$ dex.

\end{abstract}




\section{Introduction}
\label{s:introduction}

Supermassive black holes (SMBHs) are believed to exist in the centres of most, if not all galaxies \citep{Kormendy1995,Magorrian1998,Ferrarese2000,Gebhardt2000,Tremaine2002,Ho2008,Gultekin2009,Kormendy2013,Heckman2014}. When SMBHs are actively accreting matter, they are called active galactic nuclei (AGN). AGNs are also called quasars when they outshine their host galaxies. With their accretion energy output, it is likely that SMBHs are capable of modulating both their own mass accretion and the star formation of their host galaxies. \citep{Silk1998,Bower2006,Somerville2008,Sijacki2015}. On the other hand, galaxies can also regulate central black hole growth via the fueling of gas and galaxy mergers. Consequently, both SMBHs and their host galaxies may influence each others' growth, also known as ``coevolution.''  Hence, to understand galaxy and SMBH assembly histories, it is critical to constrain the interaction between SMBHs and their host galaxies \citep[see, e.g.,][]{Hopkins2006,Ho2008,Alexander2012,Kormendy2013,Heckman2014,Brandt2015}.

Two lines of observations support the coevolution scenario. First, there are relatively tight scaling relations ($\sim 0.3$ dex scatter) between SMBH masses and host galaxy dynamical properties (e.g., velocity dispersion or bulge mass at $z\sim0$ \citep[see][]{Haring2004,Gultekin2009,Kormendy2013,McConnell2013,Savorgnan2016}. Second, the cosmic SMBH accretion rate (CBHAR) density is roughly proportional to cosmic star formation rate (CSFR) density over $0<z<4$, and the constant CBHAR/CSFR ratio is $\sim 10^{-4}-10^{-3}$ \citep{Merloni2004, Silverman2008, Shankar2009, Aird2010, Delvecchio2014, Yang2018}. However, it has been difficult, if not impossible, to verify other predictions of the coevolution model (e.g., tight SMBH--galaxy property relationships) at higher redshifts.

At $z=0$, high spatial resolution spectroscopy and dynamics modeling have been used to measure SMBH--galaxy scaling relations \citep[e.g.,][]{Magorrian1998,Ferrarese2005,McConnell2013}. Beyond $z=0$, it is difficult to measure individual SMBH masses in the same way, due to lower spatial resolutions. Hence, we have to rely on indirect methods such as reverberation mapping \citep{Blandford1982,Peterson2004} or ``virial estimates'' with single-epoch AGN spectra, which infer SMBH masses from spectral line width and AGN luminosity (\citealt{Vestergaard2006}).  All such indirect methods require that SMBHs are actively accreting, and thus: 1) are biased against dormant SMBHs, and 2) make it difficult to measure host galaxy masses at the same time. In addition, overmassive SMBHs are brighter and more likely to be included in the sample, when different SMBHs have similar Eddington ratios. This bias affects the measurement of the galaxy mass--SMBH mass (\bhsm{}) relation with bright quasars \citep{Lauer2007}. \citet{Zhang2023} quantified this bias as a function of AGN luminosity, and showed that AGNs with bolometric luminosities $L_\mathrm{bol}\lesssim 10^{45}$ erg/s are needed to measure the unbiased \bhsm{} relation at $z=6$.

Prior to JWST, the highest-redshift AGN ever found was at $z\sim7.6$ \citep{Wang2021}. AGN luminosity functions at up to $z\sim 7$ have also been measured \citep{Wang2018,Kulkarni2019,Shen2020}. Most $z\sim7$ AGNs were found to be bright quasars with high SMBH masses, accreting at around the Eddington rate \citep{Shen2019}. However, given the capabilities of existing telescopes, it was difficult to ascertain if such mass and Eddington ratio distributions were representative of typical high redshift SMBHs.

JWST has changed the landscape of high-redshift quasar observations. As of the writing of this paper, JWST has found at least two AGN candidates above $z\sim 8$, i.e., CEERS1019 at $z=8.7$ \citep{Larson2023} and GN-z11 at $z=10.6$ \citep{Maiolino2023}. CEERS1019 has an estimated SMBH mass $M_\bullet\sim 10^7 M_\odot$, and an Eddington ratio ($\eta$) of $\sim 1.2$, whereas GN-z11 is estimated to be a $M_\bullet\sim 10^6 M_\odot$ SMBH accreting at an Eddington ratio of $\eta \sim 5$. JWST's sensitivity finally allows measuring SMBH masses in AGNs with $L_\mathrm{bol}\lesssim 10^{45}$ erg/s at $z\gtrsim 6$ (e.g., CEERS1019 and GN-z11), as well as measuring host galaxy properties for high-z X-ray AGN candidates, like UHZ-1 at $z=10.1$ \citep{Bogdan2023,Goulding2023}. With its unique near-infrared (NIR) sensitivity, JWST will continue to provide strong tests/constraints on theoretical models of the halo--galaxy--SMBH connection. 

Different theoretical studies have been carried out to ascertain the possible origins and evolution histories of these highest-redshift AGNs. Using a semi-analytical model, the Cosmic Archaeology Tool (CAT, \citealt{Trinca2022}), \citet{Schneider2023} found a potential evolutionary connection between objects like GN-z11 and CEERS\_1019. \citet{Scholtz2023} estimated GN-z11's host halo mass to be $\sim 3\times 10^{10} M_\odot$, based on the stellar mass--halo mass relation from the empirical \textsc{Universe Machine} model of the halo--galaxy connection \citep{Behroozi2019}. This halo mass scale makes GN-z11 a candidate protocluster core, whose $z=0$ descendant could reach above $10^{15} M_\odot$.

In this paper, we present the SMBH properties at $z\gtrsim 6$ from the empirical \textsc{Trinity} model. We further put CEERS1019 and GN-z11 in the context of \textsc{Trinity}, and examine their consistency with our predictions. \textsc{Trinity} makes statistical connections between dark matter haloes, galaxies, and SMBHs from $z=0-10$, extending the empirical DM halo--galaxy model from \citet{Behroozi2013}. Specifically, \textsc{Trinity} infers SMBH growth histories as a function of halo/galaxy mass in a similar manner as applying the So\l{}tan Argument \citep{Soltan1982} to each halo/galaxy population (see also \citealt{Shankar2020}). Compared to previous empirical models, \textsc{Trinity} includes joint constraints from a large and diverse compilation of data. Specifically, the following data are used to constrain the halo--galaxy connection: galaxy stellar mass functions (SMFs), galaxy UV luminosity functions (UVLFs), galaxy quenched fractions (QFs), galaxy specific star formation rates (SSFRs), and cosmic star formation rates (CSFRs). Additionally, the galaxy--SMBH connection is constrained by quasar luminosity functions (QLFs), quasar probability distribution functions (QPDFs), active black hole mass functions (ABHMFs), and the $z=0$ bulge mass--black hole mass relations. To accommodate the  enormous joint constraining power of this dataset, \textsc{Trinity} has a more flexible parametrization than past approaches (e.g., \citealt{Marconi2004,Shankar2009,Shankar2013,Conroy2013}. Last but not least, \textsc{Trinity} features more realistic modeling of AGN observables by including, e.g., black hole mergers and kinetic AGN luminosities in the model.

Following \citet{Behroozi2013}, \textsc{Trinity} modeling starts with population statistics from a dark matter N-body simulation. The model then generates mock universes, each with different galaxy and SMBH mass--growth rate distributions, as functions of halo mass and redshift. Millions of mock universes are and compared with real observations using a Markov Chain Monte Carlo (MCMC) algorithm to converge on the range of halo--galaxy--SMBH connections consistent with the real Universe.

The paper is organized as follows. In \S\ref{s:method}, we briefly introduce the \textsc{Trinity} framework. \S\ref{s:sims_and_data} covers the N-body simulations and galaxy/SMBH connection. For full details about the \textsc{Trinity} methodology, simulations, and datasets, we refer readers to \citet{Zhang2021}. In \S\ref{s:results}, we present high redshift predictions from \textsc{Trinity}. \S\ref{s:discussion} compares CEERS1019 and GN-z11 with \textsc{Trinity}'s predictions, and discusses how future observations would help constrain the demographics and growth histories of high redshift SMBHs. Finally, our conclusions are given in \S\ref{s:conclusions}. In this work, we adopt a flat $\Lambda$CDM cosmology with parameters ($\Omega_m=0.307$, $\Omega_{\mathrm{\Lambda}}=0.693$, $h=0.678$, $\sigma_8=0.823$, $n_s=0.96$) consistent with \textit{Planck} results \citep{Planck2016}. We use datasets that adopt the Chabrier stellar initial mass function \citep[IMF, ][]{Chabrier2003}, the \citet{Bruzual2003} stellar population synthesis model, and the Calzetti dust attenuation law \citep{Calzetti2000}. Halo masses are calculated following the virial overdensity definition from \citet{Bryan1998}.

\section{Methodology}
\label{s:method}

In \S\ref{ss:justification}, we explain the key aspects of the \textsc{Trinity} high-redshift halo--galaxy--SMBH connection that influence the predictions in this work, and how they are constrained by our data compilation. We then give a brief overview on \textsc{Trinity}'s implementation in \S\ref{ss:overview}. For full details, we refer readers to \S2 of \citet{Zhang2021}.

\subsection{Inferring the high-redshift halo--galaxy--SMBH connection with \textsc{Trinity}}
\label{ss:justification}
As an empirical model, \textsc{Trinity} infers the range of halo--galaxy--SMBH connections that match our unique compilation of galaxy$+$AGN data. Many predictions in this work come from the following two aspects of this connection:

Firstly, the intrinsic \bhsm{} relation evolves only mildly with redshift (see \S\ref{ss:bhsm_bright_quasars}). The largest change occurs for low-mass galaxies (i.e., $M_*\sim 5\times 10^{9} M_\odot$) which have lower SMBH masses at higher redshifts; almost no change in SMBH masses with redshift occurs for larger galaxies with masses $M_*\gtrsim 5\times 10^{10} M_\odot$. This redshift evolution is mainly constrained by the combination of QLFs, QPDFs, and the local \bhbm{} relation. Specifically, different redshift evolutions of the \bhsm{} relation produce different average SMBH growth rates as functions of host galaxy mass and redshift. Therefore, there is only one unique combination of a redshift-dependent \bhsm{} relation and an AGN energy efficiency that can simultaneously reproduce: 1) the probability distribution of AGN luminosity as a function of galaxy mass and redshift, i.e., QPDFs; 2) QLFs; and 3) the local \bhbm{} relation. For example, a steeper \bhsm{} relation will not only violate the constraint from the local \bhbm{} relation, but also induce faster growth of SMBHs at fixed galaxy mass, and thus overproduce AGN activity compared to observed QPDFs.

Secondly, the AGN Eddington ratio distribution becomes narrower towards higher redshifts, along with increasing average Eddington ratios. This is required by the ABHMFs from \citet{Kelly2013}. With the redshift-dependent \bhsm{} relation and AGN duty cycles constrained by QPDFs, QLFs, and the local \bhbm{} relation, lower Eddington ratios and/or broader Eddington ratio distributions will underpredict the number of SMBHs above certain luminosity limits, i.e., ABHMFs, at higher redshifts.

Of note, the AGN data for constraining \textsc{Trinity} spans a redshift range of $z=0-6.5$. When making predictions for SMBHs beyond $z\sim 7$, we are effectively examining the observational consequences of a relatively non-evolving SMBH--galaxy relationship at those redshifts.  Hence, a failure of \textsc{Trinity} to predict $z>6$ observations would indicate different evolution in the SMBH--galaxy mass relationship than that seen at $z<6$.  As we will show in \S\ref{ss:bhsm_bright_quasars} and \S\ref{ss:discussion_z9_z11_agns}, the SMBH properties of CEERS1019 and GN-z11 are partly consistent with these extrapolations. The partial inconsistency between \textsc{Trinity} and these two AGNs does not impose strong constraints on the current \textsc{Trinity} model--i.e., the inconsistency is not yet very statistically significant. As a result, further observations of high-redshift AGNs are needed to either confirm our prediction, or otherwise provide stronger constraints on the \textsc{Trinity} model.

\subsection{Implementation overview}
\label{ss:overview}

\textsc{Trinity} makes self-consisitent and statistical halo--galaxy--SMBH connections from $z=0-10$. Instead of following individual haloes across cosmic time like the \textsc{Universe Machine} \citep{Behroozi2019}, \textsc{Trinity} is built upon halo statistics from N-body simulations. The halo--galaxy connection is made via the $\mathrm{SFR}$--$M_\mathrm{peak}$ relation, where $M_\mathrm{peak}$ is the peak halo mass of haloes throughout their assembly histories. With a given redshift-dependent $\mathrm{SFR}$--$M_\mathrm{peak}$ relation, \textsc{Trinity} calculates the average star formation histories for different halo populations, and integrates them over time to obtain average galaxy stellar masses ($M_*$) as functions of redshift and halo mass. Average galaxy total stellar masses are then converted into average bulge masses ($M_\mathrm{bulge}$) with a redshift-dependent but fixed $M_\mathrm{bulge}$--$M_*$ based on SDSS \citep{Mendel2014} and CANDELS \citep{Lang2014} observations. We parametrize the \bhbm{} relation as a redshift-dependent power-law for \emph{all} galaxies, i.e., SMBH hosts and non-hosts. The log-normal scatter around the \bhbm{} relation, $\sigma_\mathrm{BH}$, is a free parameter to be constrained in MCMC. We assume that $\sigma_\mathrm{BH}$ is redshift-independent, due to the lack of data to constrain its redshift dependence. As shown in \citet{Zhang2021}, our results do not change significantly if we instead parameterize the \bhsm{} relation as a power-law. SMBH occupation fractions are parameterized as a sigmoid function of halo mass, which evolves with redshift.  Together, these choices fully parameterize the SMBH mass distribution as a function of galaxy mass and redshift ($P(M_\bullet|M_\ast,z)$). Average black hole accretion rates (BHARs) are calculated by comparing the average \mbh{} of the same halo population between two consecutive snapshots. To forward model SMBH observables, we convert average BHARs into AGN Eddington ratio distributions using parametrized AGN efficiency, duty cycles, and Eddington ratio distribution shapes (i.e., $P(L_\mathrm{bol}|M_\bullet,M_\ast,z)$). Before comparing predicted observables with real data, we also correct them for systematic effects like errors in stellar masses and SFRs from SED fitting, and Compton-thick AGN obscuration. To get the posterior distribution of model parameters, we compare predicted observables with the compiled data for $\sim2$ million points in parameter space with a custom made Metropolis Markov Chain Monte Carlo (MCMC) algorithm (based on \citealt{Haario2001}). From this posterior distribution, we obtain both the best-fitting model parameter set given the observational constraints, as well as the uncertainties in the halo--galaxy--SMBH connection and AGN properties, e.g., efficiency and duty cycles. The whole parametrization of \textsc{Trinity} contains 54 free parameters.

\subsection{Calculating SMBH statistics}
\label{ss:smbh_stats}
In this work, we present various statistics of SMBHs at $z\gtrsim 6$ from the best-fitting \textsc{Trinity} model. Here, we briefly introduce how we calculate them.

To calculate the median \bhsm{} relation for bright quasars above a certain bolometric luminosity threshold ($L_{\mathrm{bol,lim}}$), we firstly calculate the SMBH mass function at each given stellar mass and redshift:

\begin{equation}
\label{e:bhmf_quasars}
\begin{aligned}
\phi_{\bullet}\left(M_\bullet | M_*, L_{\mathrm{bol}} > L_{\mathrm{bol,lim}}, z\right) = &\int_0^\infty \phi_{\mathrm{h}}\left(M_{\mathrm{peak}}, z\right) \\ 
&\int_{L_{\mathrm{bol,lim}}}^\infty  P(M_\bullet | M_*, z) P(M_* | M_\mathrm{peak}, z)\\
&P\left(L_{\rm bol}|M_{\rm peak}, z\right)dL_{\mathrm{bol}}d M_{\mathrm{peak}}\ ,
\end{aligned}
\end{equation}
where $P\left(L_{\rm bol}|M_{\rm h}, z\right)$ is obtained by integrating the number density of black holes emitting at the corresponding Eddington ratio:

\begin{equation}
\begin{aligned}
P\left(L_{\rm bol}|M_{\rm peak}, z\right) = \int_0^\infty &P\left(\eta_{\rm rad}\left(L_{\rm bol}, M_{\bullet}\right)|M_{\bullet},M_{\rm peak},z\right)\times\\
&P\left(M_{\bullet}|M_{\rm peak},z\right)d M_{\bullet}\label{e:qlf_2}\ .
\end{aligned}
\end{equation}
We then calculate the median \mbh{} and the 16$^\mathrm{th}$--84$^\mathrm{th}$ percentile range from $\phi_{\bullet}\left(M_\bullet | M_*, L_{\mathrm{bol}} > L_{\mathrm{bol,lim}}, z\right)$. The \bhsm{} relation for bright quasars is then obtained by repeating this process for a series of \mstar{} at fixed $z$.

Similarly, SMBH mass functions are convolutions of halo mass functions and the conditional probability distribution of \mbh{} given \mpeak{}:
\begin{equation}
\label{e:bhmf}
\begin{aligned}
\phi\left(M_\bullet, z\right) &= \int_0^\infty \phi_{\mathrm{h}}\left(M_{\mathrm{peak}}, z\right)P\left(M_\bullet|M_{\rm peak}, z\right)d M_{\mathrm{peak}}\ ,
\end{aligned}
\end{equation}
where $P\left(M_\bullet|M_{\rm peak}, z\right)$ is the convolution of: 1) the probability distribution of $M_\bullet$ given \mstar{}, $P(M_\bullet|M_*,z)$; and 2) the probability distribution of \mbh{} given \mpeak{}, $P(M_*|M_\mathrm{peak},z)$:
\begin{equation}
\label{e:prob_mbh_mh}
\begin{aligned}
P\left(M_\bullet|M_{\rm peak}, z\right) &= \int_0^\infty P(M_\bullet|M_*,z) P(M_*|M_\mathrm{peak},z) d M_*\ .
\end{aligned}
\end{equation}

The cosmic number density of SMBHs above a mass threshold $M_{\bullet,\mathrm{lim}}$, $n_\bullet(M_\bullet>M_{\bullet,\mathrm{lim}}, z)$, is a convolution of $P\left(M_\bullet|M_{\rm peak}, z\right)$ and the halo mass function $\phi_\mathrm{h}(M_{\rm peak})$:
\begin{equation}
    n_\bullet(M_\bullet > M_{\bullet,\mathrm{lim}}, z) = \int_{M_{\bullet,\mathrm{lim}}}^{\infty}P\left(M_\bullet|M_{\rm peak}, z\right)\phi_\mathrm{h}(M_\mathrm{peak},z)dM_\mathrm{peak}\ .
\end{equation}
The differential number of SMBHs above $M_{\bullet,\mathrm{lim}}$ \emph{per unit redshift}, $dN_\bullet (M_\bullet > M_{\bullet,\mathrm{lim}},z)/dz$, is then:
\begin{equation}
     \frac{dN_\bullet}{dz}(M_\bullet > M_{\bullet,\mathrm{lim}},z) = n_\bullet(M_\bullet > M_{\bullet,\mathrm{lim}}, z) \times \frac{dV}{dz}\ ,
\end{equation}
where $dV/dz$ is the differential comoving volume per unit redshift, calculated based on the adopted cosmology (see \S\ref{s:introduction}). The cumulative number of SMBHs above $M_{\bullet,\mathrm{lim}}$ at above redshift $z$, $N_\bullet (M_\bullet > M_{\bullet,\mathrm{lim}},>z)$ is thus:
\begin{equation}
     N_\bullet (M_\bullet > M_{\bullet,\mathrm{lim}},>z) = \int_{z}^{\infty}\frac{dN_\bullet}{dz}(M_\bullet > M_{\bullet,\mathrm{lim}},z') dz'\ .
\end{equation}

\section{Simulations and Data Constraints}
\label{s:sims_and_data}

\subsection{Dark Matter Halo Statistics}
\label{ss:dm_sims}

As noted in \S \ref{ss:overview}, \textsc{Trinity} requires only halo population statistics from dark matter simulations, as opposed to individual halo merger trees.  We use the peak historical mass (\mpeak{}) halo mass functions for central and satellite haloes from \citet{Behroozi2013}, for the cosmology specified in the introduction. These mass functions are based on central halo mass functions from \citet{Tinker2008}, with adjustments to include satellite halo number densities as well as to use \mpeak{} instead of the present day halo mass. These adjustments were based on the Bolshoi \& Consuelo simulations \citep{Klypin2011}. We refer readers to Appendix G of \citet{Behroozi2013} for full details. With these calibrations, the halo statistics used in this work are suitable for studying the evolution of halos from $10^{10} M_\odot$ to $10^{15} M_\odot$. 

Haloes grow by smooth accretion and halo mergers. For average halo mass accretion histories, we use the fitting formulae in Appendix H of \citet{Behroozi2013}. For halo mergers, we fit merger rates from the \textsc{UniverseMachine} \citep{Behroozi2019}, with full details and formulae in Appendix B of \citet{Zhang2021}.

In addition to statistical halo assembly histories, we also use individual haloes from the MultiDark Planck 2 simulation (MDPL2, \citealt{Klypin2016}) to study the evolution of $z=6$ haloes hosting SMBHs with $M_\bullet > 10^{9} M_\odot$. The MDPL2 simulation assumes Planck cosmology \citep[][]{Planck2016}, and has a simulation box size of 1 Gpc$/h$. The mass of each dark matter particle is $1.51\times 10^9 M_\odot/h$, which enables the simulation to resolve haloes down to $M_\mathrm{peak}\sim 10^{11} M_\odot$. As we will show in this work, the combination of the mass resolution and simulation box volume makes MDPL2 suitable for studying the growth histories of haloes hosting SMBHs with $M_\bullet > 10^{9} M_\odot$ at $z\sim 6$, as well as the descendants of typical haloes hosting GNz-11 and CEERS\_1019.

\subsection{Observational Data Constraints}
\label{ss:obs_data}

\label{sss:obs_data_original}

To constrain the 54 free parameters in \textsc{Trinity}, we have originally compiled a diverse dataset including galaxy and SMBH observables. The galaxy data include: galaxy stellar mass functions (SMFs), quenched fractions (QFs), cosmic star formation rates (CSFRs), average specific star formation rates (SSFRs), and UV luminosity functions (UVLFs). Collectively, these data cover a redshift range of $0\leq z\leq 10$. The SMBH observables are: quasar luminosity functions (QLFs), quasar probability distribution functions (QPDFs), active black hole mass functions (ABHMFs), the $z=0$ \bhbm{} relation, and the observed \mbh{} distribution of high redshift bright quasars. These SMBH observables collectively cover a redshift range of $0\leq z\leq 6.5$. These datasets are summarized in Tables 3-10 of \citet{Zhang2021}, and we refer readers to Appendix C of \citet{Behroozi2019} and Appendix D of \citet{Zhang2021} for the details about the homogenization of galaxy and AGN data, respectively.

\label{sss:obs_data_new}

\begin{figure}
\includegraphics[width=0.48\textwidth]{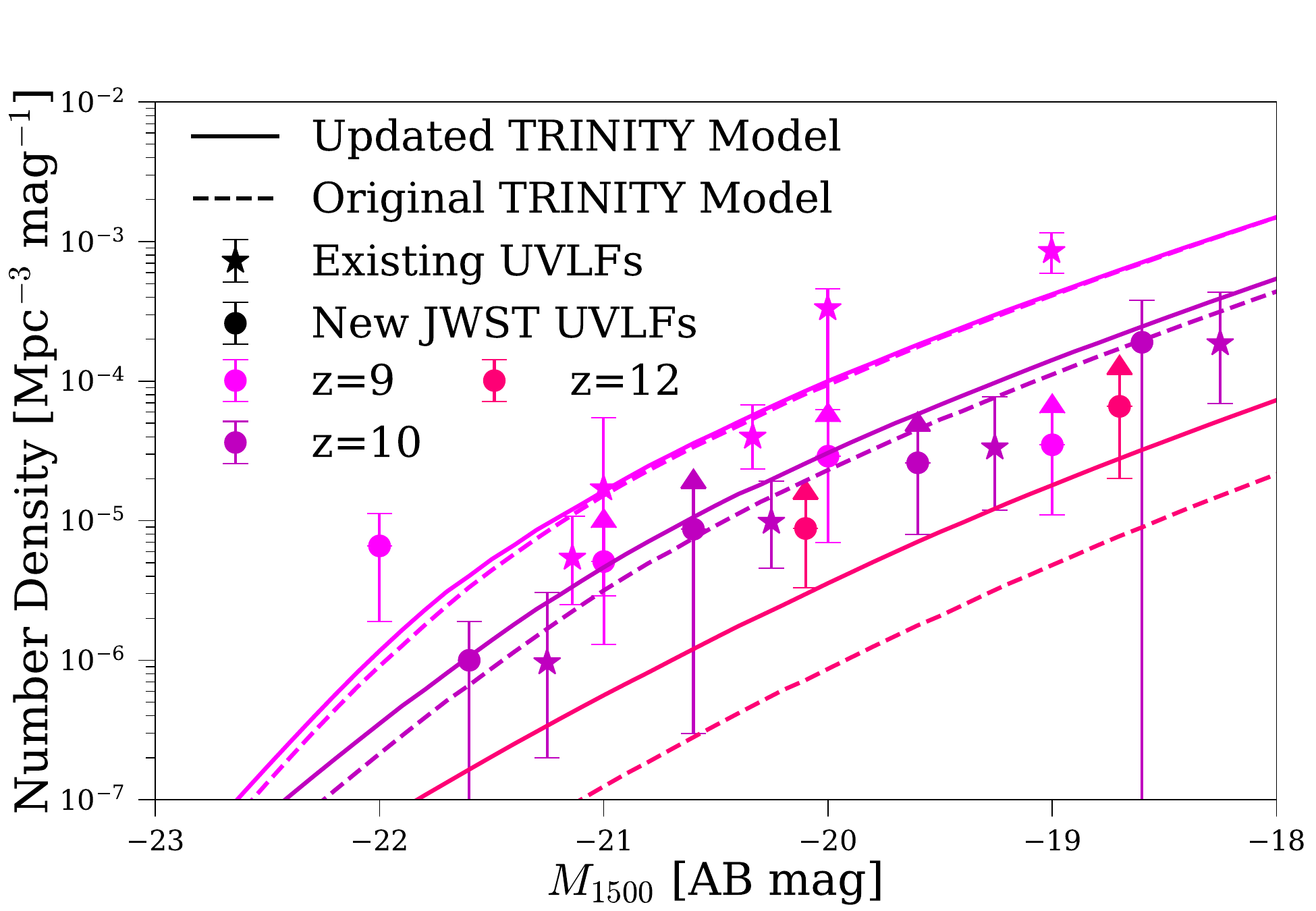}
\caption{The comparison between the original \textsc{Trinity} model (dashed lines, \citealt{Zhang2021}), the updated \textsc{Trinity} model (solid lines), the existing galaxy UVLF data (stars, \citealt{Ishigaki2018,Oesch2018,Bouwens2019}, also see Table 8 of \citealt{Zhang2021}), and the new galaxy UVLFs from JWST \citep{Harikane2023}. The difference between the original and updated \textsc{Trinity} models increases with redshift. See \S\ref{sss:obs_data_new}.}
\label{f:jwst_uvlf}
\end{figure}

Since its launch, JWST has found more high-redshift UV-bright galaxies than predicted by theoretical models such as the \textsc{UniverseMachine}. A higher number density of high-redshift galaxies likely implies more high-redshift SMBHs, which would systematically change our prediction for the early Universe. We thus added the $9 \lesssim z \lesssim 13$ galaxy UVLFs from \citet{Harikane2023b} to our previous data compilation, which includes only galaxies confirmed with spectroscopy. In \textsc{Trinity}, we calculate galaxy UV magnitudes with a scaling relation between UV magnitudes and SFR, to get statistically equivalent results to the Flexible Stellar Population Synthesis code (FSPS, \citealt{Conroy2013}). For full details of calculating galaxy UV magnitudes, we refer the readers to Appendix B of \citet{Zhang2021}.

The addition of $9 \lesssim z \lesssim 13$ galaxy UVLFs does not induce any qualitative changes in the results presented by \citet{Zhang2021}, \citet{Zhang2023}, or \citet{Zhang2023c}. Quantitatively, it does increase the number of UV-bright galaxies at $z\gtrsim 9$, and ultimately, the galaxy masses at fixed halo masses. This increase in galaxy number density and typical galaxy mass also leads to more SMBHs at fixed SMBH masses, because the redshift-dependent \bhsm{} relation is unchanged with the addition of new galaxy UVLFs. As shown in Fig.\ \ref{f:jwst_uvlf}, this change in galaxy number density increases towards higher redshifts, especially above $z=10$. This is because the new JWST UVLFs (solid circles) are largely consistent with existing UVLFs in our data compilation (solid stars), but we lacked constraints at $z\sim 12$ in the original \textsc{Trinity} model. Without $z\sim 12$ constraints, the original \textsc{Trinity} underproduced UV-bright galaxies, leading to the larger difference at this redshift. The results presented in this work are based on the \emph{updated} \textsc{Trinity} model after adding these $9 \lesssim z \lesssim 13$ galaxy UVLFs to our data constraints. We have also updated the plots in previous \textsc{Trinity} papers \href{https://github.com/HaowenZhang/TRINITY}{\textbf{here}}.

\section{Results}
\label{s:results}

We present the offset between the $z=6-10$ \bhsm{} relations for high redshift bright quasars vs.\ all SMBHs in \S\ref{ss:bhsm_bright_quasars}, and the cosmic abundance of very massive black holes ($M_\bullet > 10^9 M_\odot$ and $10^{10} M_\odot$) in \S\ref{ss:results_huge_bh}. In \S\ref{ss:desc_big_mbh_at_z6}, we show the $z=0$ descendant mass distribution of the $z=6$ haloes hosting $M_\bullet > 10^{9} M_\odot$ and $10^{10} M_\odot$ SMBHs, and discuss the prospect of detecting these descendants in the local Universe.

\subsection{The biased \bhsm{} relation for bright quasars between $6\lesssim z \lesssim 11$}
\label{ss:bhsm_bright_quasars}

\begin{figure}
\vspace{-1cm}
\includegraphics[width=0.48\textwidth]{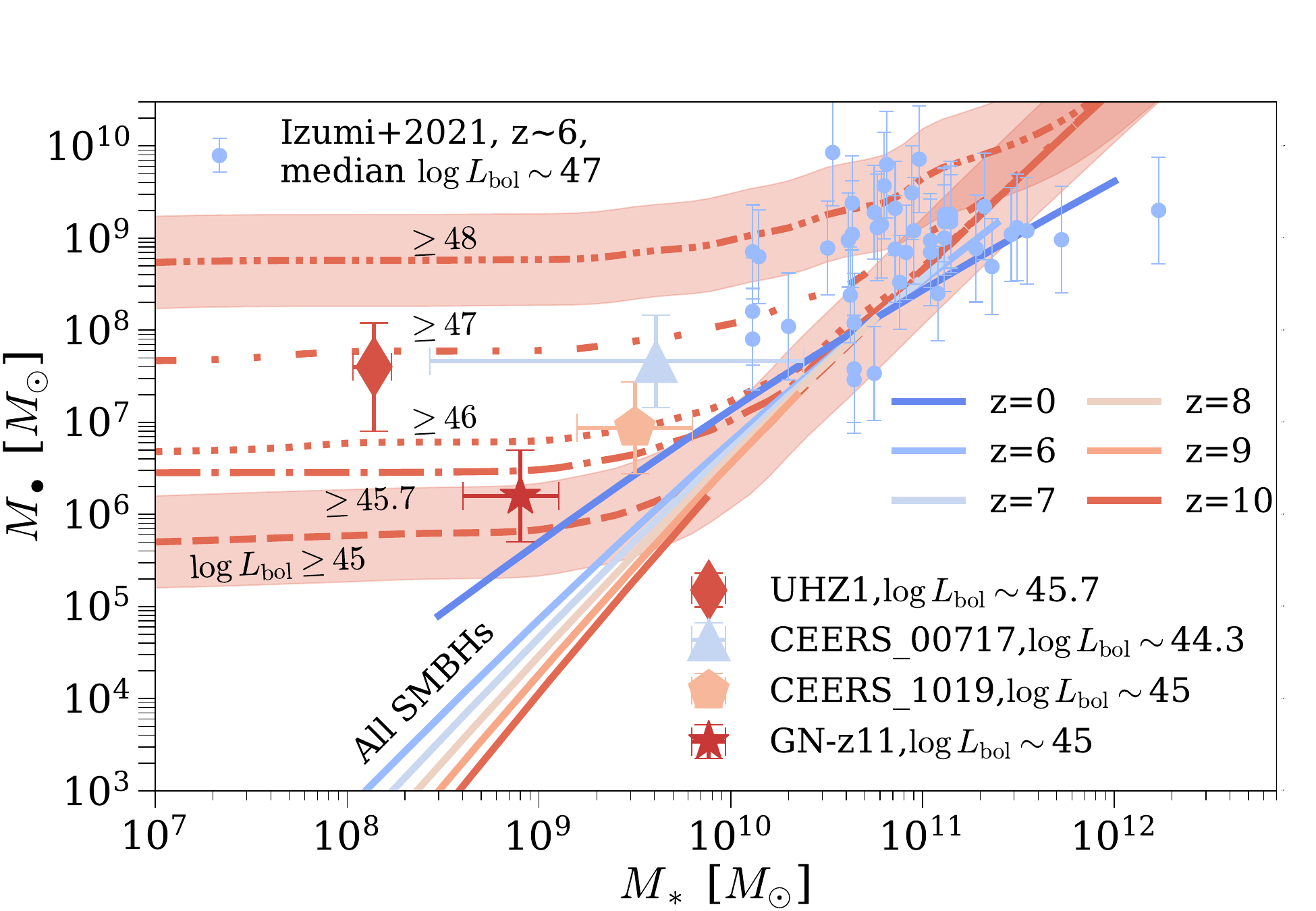}
\caption{The observed median \bhsm{} relations for bright quasars selected by different bolometric luminosity thresholds at $z=6-11$. The dashed, dash-dotted, dotted, and dash-double-dotted curves are the \bhsm{} relations for bright quasars above different luminosity limits, which are annotated next to each line. These \bhsm{} relations are nearly redshift-independent, so we only show the ones at $z=10$ for visual clarity. The shaded regions are the $1-\sigma$ log-normal scatter ($\sim 0.6$ dex) around the \bhsm{} relations for bright quasars, which includes the intrinsic scatter around the \bhsm{} relation ($\sim 0.3$ dex), as well as the typical random scatter in \mbh{} from virial estimates ($\sim 0.5$ dex). Since the magnitude of this scatter is nearly luminosity-independent, we only show it for the brightest and faintest quasars, for clarity. The solid lines of different colours represent the intrinsic median \bhsm{} relations for \emph{all SMBHs} at each redshift. UHZ1 \citep{Bogdan2023,Goulding2023}, CEERS\_00717 \citep{Harikane2023}, CEERS\_1019 \citep{Larson2023} and GN-z11 \citep{Maiolino2023} are shown in the diamond, the triangle, the pentagon, and the star, respectively. $M_\bullet$ uncertainties of 0.5 dex are assumed for CEERS\_00717, CEERS\_1019, and GN-z11. For UHZ1, the upper uncertainty of $M_\bullet$ is assumed to be 0.5 dex, but the lower uncertainty corresponds to changing the assumed Eddington ratio from 1 to 5, i.e., 0.7 dex. The galaxy mass uncertainties for these four high-redshift AGNs are taken from the source papers. We also show the $z\sim6$ bright quasars with median $\log L_\mathrm{bol}\sim 47$ compiled by \citet{Izumi2021} in solid circles with error bars ($\sim 0.5$ dex). See \S\ref{ss:bhsm_bright_quasars}.}
\label{f:bhsm_bias}
\end{figure}

In \citet{Zhang2023}, we showed the median \bhsm{} relation for bright quasars at $z\sim 6$, and concluded that overmassive SMBHs compared to the observed local \bhsm{} relation are consistent with arising purely from selection bias against less massive and thus fainter quasars. In Fig.\ \ref{f:bhsm_bias}, we extend our predictions to even higher redshifts, i.e., $6\lesssim z \lesssim 11$. The solid lines denote the intrinsic (i.e., including active and inactive SMBHs) median \bhsm{} relation at different redshifts. The non-solid curves are the \bhsm{} relations for quasars above various bolometric luminosity limits, which are labeled next to each line. Given their weak redshift dependence, we only show these quasar \bhsm{} relations at $z=11$ for visual clarity. The shaded regions correspond to $1-\sigma$ log-normal scatter of $\sim 0.6$ dex around these quasar \bhsm{} relations, which includes: 1) the intrinsic scatter around the \bhsm{} relation ($\sim 0.3$ dex, according to the best-fit \textsc{Trinity} model); and 2) the random scatter around \mbh{} from virial estimates (0.5 dex, see, e.g., \citealt{McLure2004,Vestergaard2009}). The scatter is nearly luminosity-independent, so we only show the scatter for the brightest quasars with $\log L_\mathrm{bol} > 48$ and the deepest samples with $\log L_\mathrm{bol} > 45$, also for clarity. 

Similar to those at $z\sim 6$, the \bhsm{} relations for bright quasars continue to be higher than the intrinsic relation for all SMBHs. This is because at high redshifts, different SMBHs share very similar Eddington ratio distributions (see \S\ref{ss:justification}). When selected by bolometric luminosity, more massive SMBHs are brighter and naturally over-represented in the sample. At fixed bolometric luminosity limits, the \bhsm{} relations for quasars barely evolve with redshift, despite the evolution in the \emph{intrinsic} \bhsm{} relation for \emph{all} SMBHs. This is because at fixed SMBH mass, the Eddington ratio distribution evolves very mildly beyond $z\sim 6$. Thus, selecting quasars by their luminosities yields similar SMBH masses across different redshifts, which are more and more overmassive SMBHs compared to the lower intrinsic \bhsm{} relations towards higher redshifts. 

We also note that the typical $M_\bullet$ for quasars depends only weakly on host galaxy mass below $\log M_*\sim 9.5$. This is because, on average (i.e., including their active and dormant phases), high-redshift SMBHs grow at slightly below the Eddington rate, and hyper-Eddington AGNs are very rare. This upper limit in Eddington ratios translates into a lower limit in the typical $M_\bullet$, regardless of host galaxy mass. For brighter quasars (e.g., $\log L_\mathrm{bol} \geq 48$), the typical SMBH mass remains above $3\times 10^8 M_\odot$ even for galaxies below the same mass. This implies that if we detected any $\log L_\mathrm{bol} \geq 48$ quasars in a $M_* < 3\times 10^8 M_\odot$ galaxy, they either: 1) have Eddington ratios much higher than predicted by \textsc{Trinity}; and/or 2) they are in the outsized black hole galaxy (OBG) phase \citep{Agarwal2013,Natarajan2017}.

Of note, when quantifying this luminosity dependence of the \bhsm{} relation (i.e., the Lauer bias), we assume that \emph{every} AGN above those luminosity limits is included in the samples, with SMBH mass measurements via virial estimates. In other words, we do not account for the possibility that these AGNs may be hidden in the light from host galaxies. This assumption is more likely to hold for brighter quasars. For fainter AGNs, e.g., $\log L_\mathrm{bol} \sim 45$, stronger contamination from the host galaxies makes it more difficult to measure SMBH masses, even if the AGN luminosities are technically above the limits of current instruments. As a result, overmassive SMBHs would still be preferred among fainter AGNs (see, e.g., \citealt{Volonteri2023}). It takes detailed modeling of high-redshift galaxy and SMBH spectral energy distributions (SEDs) to properly account for this bias, which we defer to a future work. Therefore, our predicted \bhsm{} relations for fainter AGNs (e.g., with $\log L_\mathrm{bol} > 45$) should be regarded as \emph{lower limits}.

We also show the four new highest-redshift AGNs detected by JWST, CEERS\_00717 (\citealt{Harikane2023}, the triangle), CEERS1019 (\citealt{Larson2023}, the pentagon), UHZ1 (\citealt{Bogdan2023,Goulding2023}, the diamond) and GN-z11 (\citealt{Maiolino2023}, the star) in Fig.\ \ref{f:bhsm_bias}. Since the SMBH masses of CEERS\_00717, CEERS\_1019, and GN-z11 are estimated with virial estimates, we adopt an $M_\bullet$ uncertainty of 0.5 dex for these three AGNs. The $M_\bullet$ of UHZ1 was estimated from X-ray spectral information, modeling jointly luminosity and obscuring column density, and further assuming Eddington-limited accretion. Given that SMBHs accreting at five times the Eddington rate have been detected at similar redshifts (i.e., GNz-11), we adopt a higher $M_\bullet$ uncertainty of 0.7 dex (i.e., a factor of five) to reflect the uncertainty in Eddington ratio. Galaxy mass uncertainties are directly taken from these source papers. We note that the stellar masses have in several cases been determined without including an AGN component in the fit, and that the uncertainties in the star formation histories can give large systematic uncertainties (see, e.g., the summary in Table 1 of \citealt{Volonteri2023}).

Compared to the brighter quasars with median $\log L_\mathrm{bol} \sim 47$ compiled by \citet{Izumi2021}, all these JWST AGNs have lower SMBH masses. This is consistent with \textsc{Trinity}'s prediction for the luminosity-dependent bias (i.e., the Lauer bias) of the \bhsm{} relation. At face value, the SMBH in CEERS\_00717 seems much more massive than \textsc{Trinity}'s prediction for faint AGNs. But given the substantial uncertainty in galaxy mass, we decide to compare our predictions with CEERS\_00717 after more precise measurements of galaxy mass are available. We do note that both CEERS1019 and GN-z11 lie above our \bhsm{} relation for AGNs with $\log L_\mathrm{bol} > 45$, at $\sim 1.5\sigma$ and $\sim0.8\sigma$, respectively. This could be due to: 1) small number statistics; 2) the aforementioned selection bias due to smaller SMBHs being harder to detect against the host galaxy's light; and/or 3) \textsc{Trinity} underpredicting SMBH masses compared to the real Universe. Further observations and better characterizations of all selection biases are needed to ascertain the nature of this difference. In \S\ref{ss:discussion_z9_z11_agns}, we compare CEERS1019 and GN-z11 with other $6\lesssim z\lesssim 11$ predictions from \textsc{Trinity}.

Finally, we make a few remarks about UHZ1. Assuming Eddington accretion, the $M_\bullet$ of UHZ1 is more than 1 dex higher than expected for AGNs above similar luminosities ($\log L_\mathrm{bol}\geq 45.7$, the dash-dotted line). But, at $z\gtrsim 10$, another super-Eddington AGN has been found, i.e., GN-z11. If UHZ1 is in fact also accreting at a similar Eddington ratio, the resulting $M_\bullet$ would be marginally consistent with \textsc{Trinity} predictions. A caveat of this argument is that the current estimate of UHZ1's luminosity (and thus $M_\bullet$ as well) is only a $1\sigma$ lower limit from X-ray spectral fitting. Such values were reported instead of the actual best-fitting values due to the strong degeneracy between the intrinsic X-ray luminosity and gas column density used in the fitting process. Thus, choosing higher intrinsic luminosity values will push UHZ1 further away from \textsc{Trinity}'s prediction. This would imply that UHZ1 may be experiencing an OBG phase, where the SMBH is not evolving in sync with the host galaxy and could become more massive than the galaxy. If confirmed, such AGNs in OBGs will provide critical information for future versions of \textsc{Trinity} so that it can include these extreme objects.

\subsection{Very massive black holes at $z>6$}
\label{ss:results_huge_bh}

\begin{figure}
\subfigure{
\includegraphics[width=0.48\textwidth]{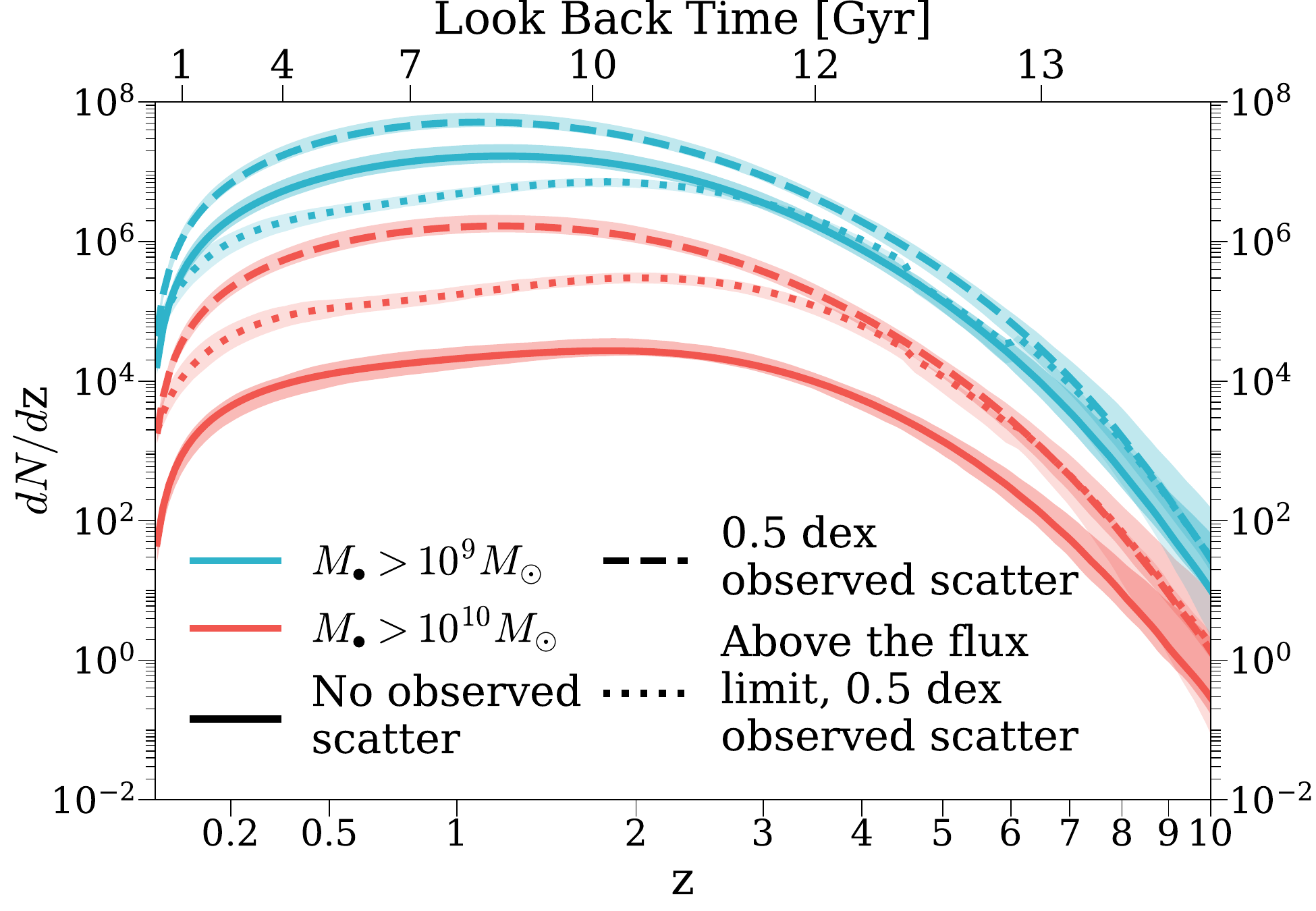}
}
\subfigure{
\includegraphics[width=0.48\textwidth]{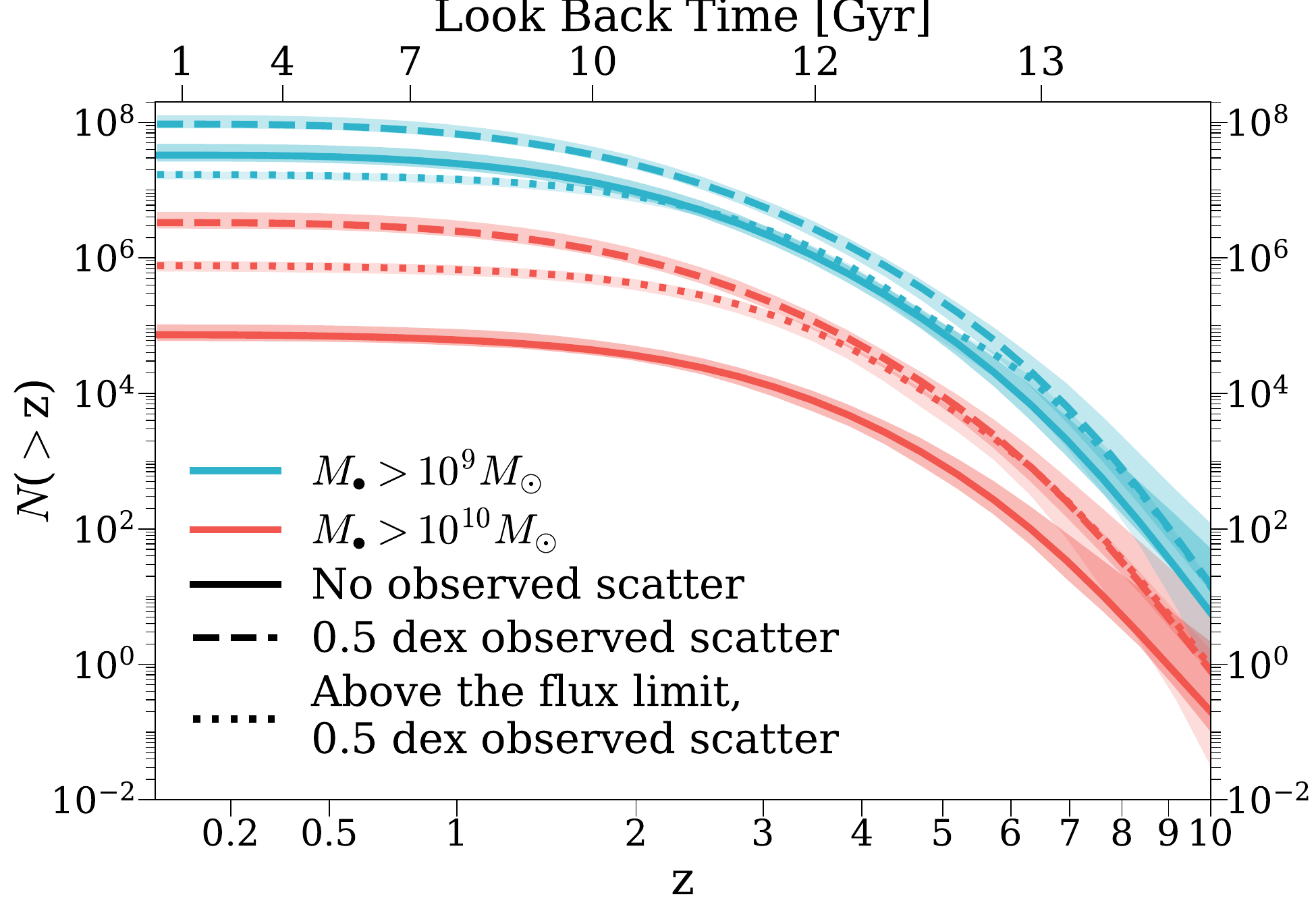}
}
\subfigure{
\includegraphics[width=0.48\textwidth]{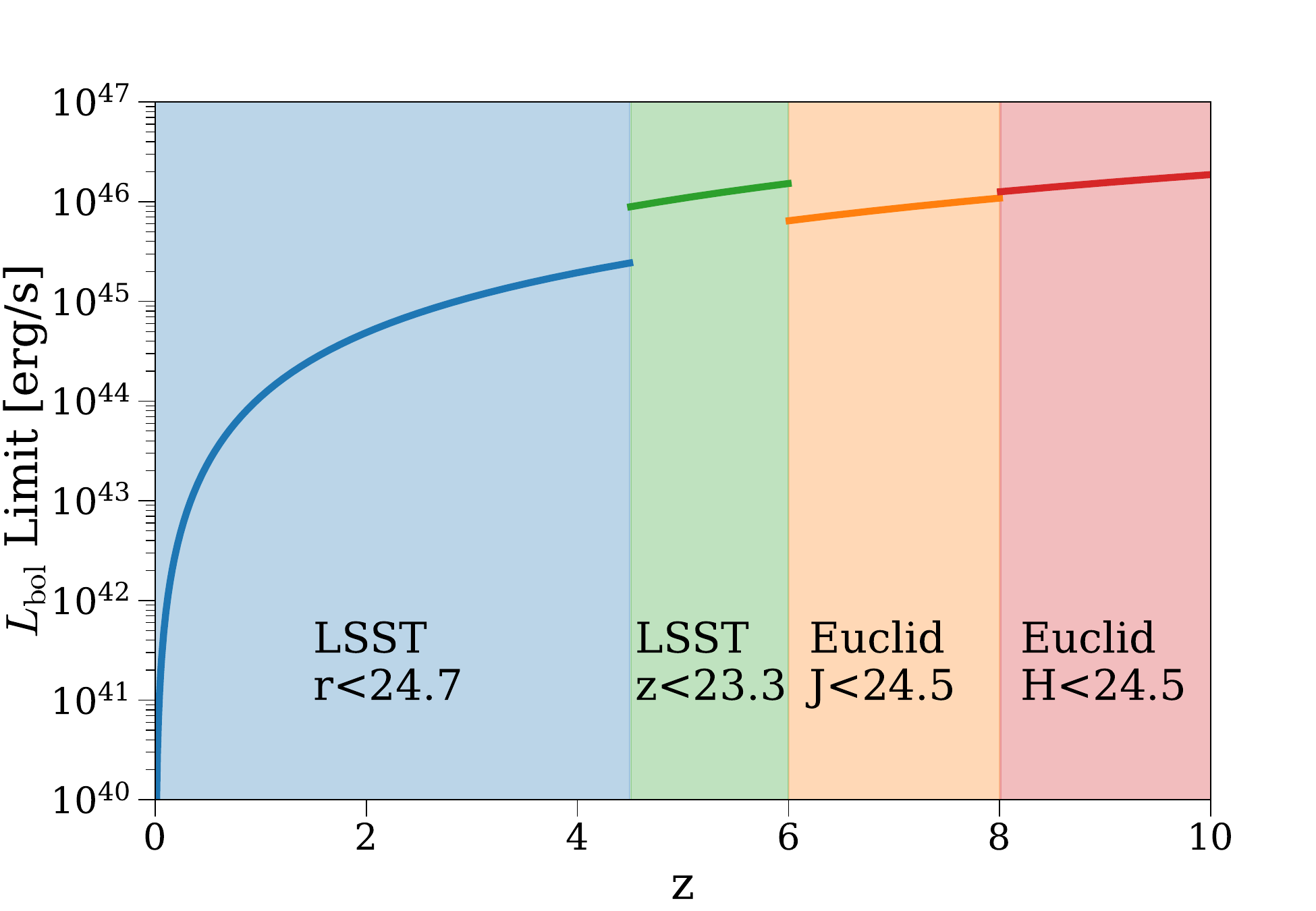}
}
\caption{\textbf{Top Panel:} The number of very massive black holes \emph{per unit redshift} in the whole observable universe (i.e., $dN/dz$). The solid lines show results assuming no observational scatter in the measured black hole masses, and the dashed lines show results assuming an observational scatter of $\sigma_{\rm vir}=0.5$ dex, typical for virial estimates (see \S\ref{ss:results_huge_bh} for details). The dotted lines show the numbers of SMBHs above the bolometric luminosity limit shown in the bottom panel. \textbf{Middle Panel:} The \emph{cumulative} number of very massive black holes as a function of redshift. \shadedregions{} \textbf{Bottom Panel:} The bolometric luminosity limit to produce the dotted lines in the top and middle panels, as a function of redshift. See \S\ref{ss:results_huge_bh}.}
\label{f:huge_bh}
\end{figure}

\begin{figure}
\includegraphics[width=0.48\textwidth]{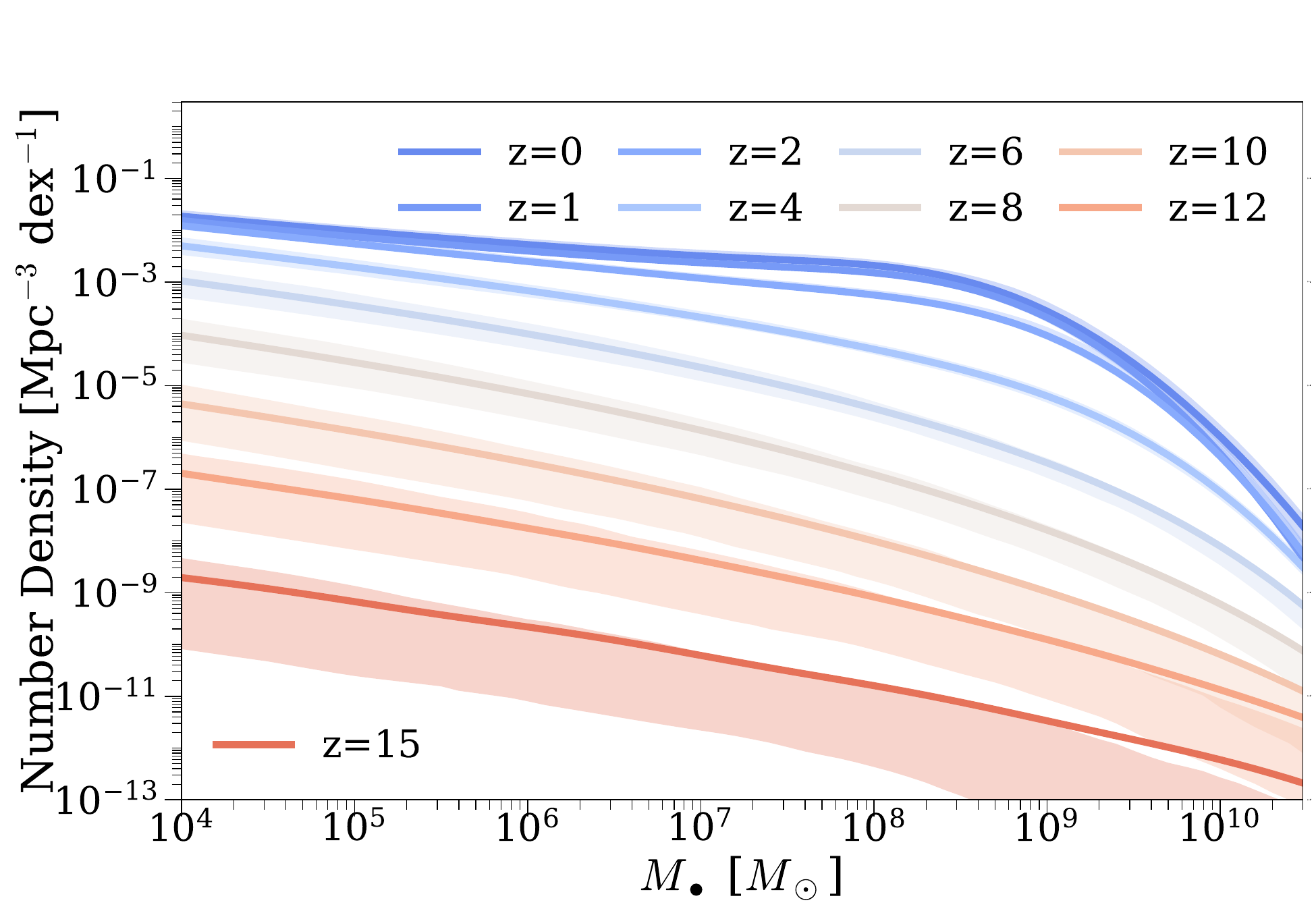}
\caption{Total SMBH mass functions from $z=0-10$. \shadedregions{} See \S\ref{ss:results_huge_bh}. This figure is reproduced from \citet{Zhang2021} to help readers understand the trends in Fig.\ \ref{f:huge_bh}.}
\label{f:bhmf}
\end{figure}

The top panel of Fig.\ \ref{f:huge_bh} shows the redshift density of very massive black holes ($M_\bullet > 10^{9} M_\odot$ and $10^{10} M_\odot$), i.e., $dN/dz$, in the whole observable Universe. The solid lines represent all (i.e., both active and dormant) SMBHs with \emph{intrinsic} masses above these thresholds. From $z=10$ to $z\sim 1-2$, $dN/dz$ increases rapidly with time due to high black hole accretion rates. Below $z\sim 1-2$, $dN/dz$ decreases due to the combination of: 1) the decreasing observable volume; 2) the decline in SMBH growth. The middle panel of Fig.\ \ref{f:huge_bh} shows cumulative numbers of very massive black holes at different redshifts in the observable Universe. Beyond the local Universe, virial estimates remain the only way to measure black hole masses for large samples. These methods result in random scatter of $0.4-0.6$ dex in observed black hole masses around the intrinsic values (\citealt{McLure2002,Vestergaard2006}). This scatter causes an Eddington bias \citep{Eddington1913}, because there are more low-mass black holes that can be upscattered than high-mass black holes that can be downscattered. This elevates the observed number of massive black holes above the true number, 
as shown by the dashed curves in the top and middle panels of Fig.\ \ref{f:huge_bh}. Overall, an observational scatter of $0.5$ dex elevates observed massive black hole number counts by a factor of $\sim 3-40$. This boost in black hole numbers is stronger for more massive objects and/or lower redshifts. This is because the massive end of the SMBH mass function becomes steeper towards lower redshifts, as is shown in Fig. \ref{f:bhmf}. The steeper mass function leads to a larger population of upscattered \mbh{} measurements, and thus a stronger boost in the numbers of large SMBHs. With this Eddington bias, the cumulative numbers of $M_\bullet > 10^{9},10^{10} M_\odot$ SMBHs increase by six orders of magnitude from $z\sim 10$ to $z\sim 2$.

According to \textsc{Trinity}, a $10^9 M_\odot$ SMBH would be $\sim 3-4$ dex above the median \bhsm{} relation at $z=10$ (the bottom right panel of Fig.\ \ref{f:bhsm_bias}). The rarity of such SMBHs is also shown in the bottom panel of Fig.\ \ref{f:huge_bh}: we expect to see only $\lesssim 10$ such SMBHs at $z\gtrsim 10$ \emph{over the whole sky, whether they are active or not}. Such rarity results from the extreme difficulty of growing a $10^9 M_\odot$ SMBH by $z\sim 10$, which requires an appropriate combination of SMBH seed mass, Eddington ratios, duty cycle, and AGN efficiency \citep{Pacucci2022}. The detection of massive BHs at $z\gtrsim 10$ will therefore provide stringent constraints on SMBH seeding mechanisms, as well as their early growth histories. We will further compare the possible progenitor masses and growth speeds of $10^9 M_\odot$ SMBHs from \citet{Pacucci2022} and \textsc{Trinity} in \S\ref{ss:compare_with_pacucci}.

To better understand the SMBH detection capability of future wide field surveys, we also made predictions of expected SMBH numbers from the \textit{Legacy Survey of Space and Time} (\textit{LSST}) and the \textit{Euclid} wide survey. To calculate the bolometric luminosity limits as a function of redshift, we adopted the magnitude limits from \citet{Ivezic2019} and \citet{Euclid2022}. We further convert observed magnitude limits in different wavebands (labeled in the bottom panel of Fig.\ \ref{f:huge_bh}) into bolometric luminosities, by assuming: 1) that the bolometric correction to the rest-frame $i$-band luminosity is 12 \citep{Richards2006}; and 2) the UV-to-optical quasar SED is a power-law with a power-law index of $\alpha_\nu = -0.44$ \citep{VandenBerk2001}. The luminosity limit as a function of redshift is shown in the bottom panel of Fig.\ \ref{f:huge_bh}, and the differential (cumulative) number of massive black holes is shown in dotted lines in the top (middle) panel of Fig.\ \ref{f:huge_bh}. 

At $z\gtrsim 4$, the vast majority of SMBHs with $M_{\bullet,\mathrm{obs}} > 10^9 M_\odot$ are active, so the bulk of these SMBHs are above the flux limits of LSST and Euclid. We note that the number of SMBHs in Fig.\ \ref{f:huge_bh} include \emph{both obscured and unobscured} SMBHs. Given the uncertainties in the obscured AGN fraction towards higher redshifts, we opt not to show results for unobscured AGNs in Fig.\ \ref{f:huge_bh}, but only give a rough estimate: at these high luminosity limits ($\gtrsim 10^{46}$ erg/s), we expect $\sim 40\%$($80\%$) of such SMBHs to be unobscured, based on the obscured fractions as functions of AGN X-ray luminosity given by \citet{Merloni2014} and \citet{Ueda2014}. Below $z\sim 4$, massive black holes experience a significant decline in activity level. Consequently, smaller and smaller fractions of massive black holes at lower redshifts will lie above the flux limit of LSST.

\subsection{Host halos of massive black holes at $z=6$ and their descendants at $z=0$}
\label{ss:desc_big_mbh_at_z6}

\begin{figure}
\subfigure{
\includegraphics[width=0.48\textwidth]{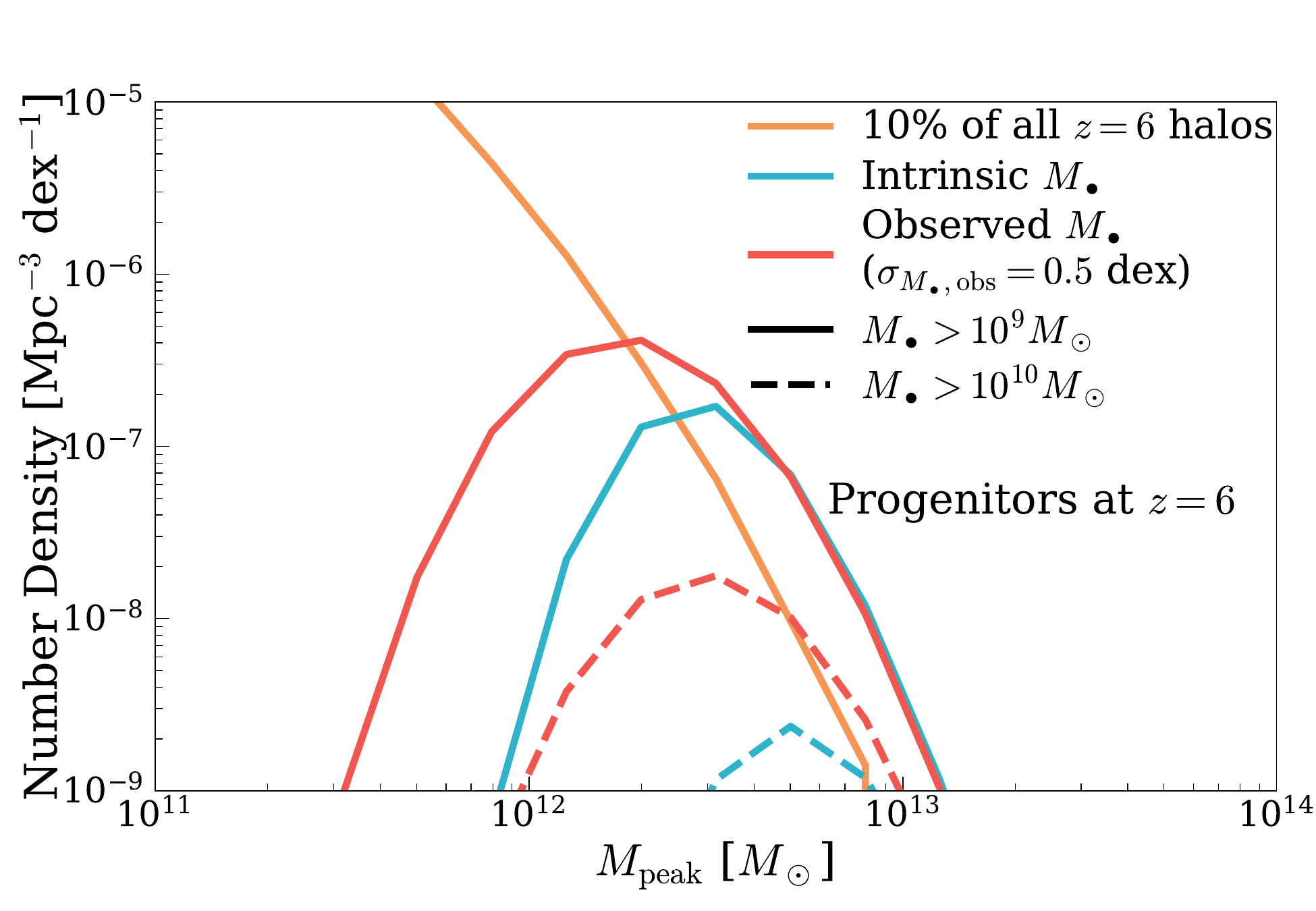}
}
\subfigure{
\includegraphics[width=0.48\textwidth]{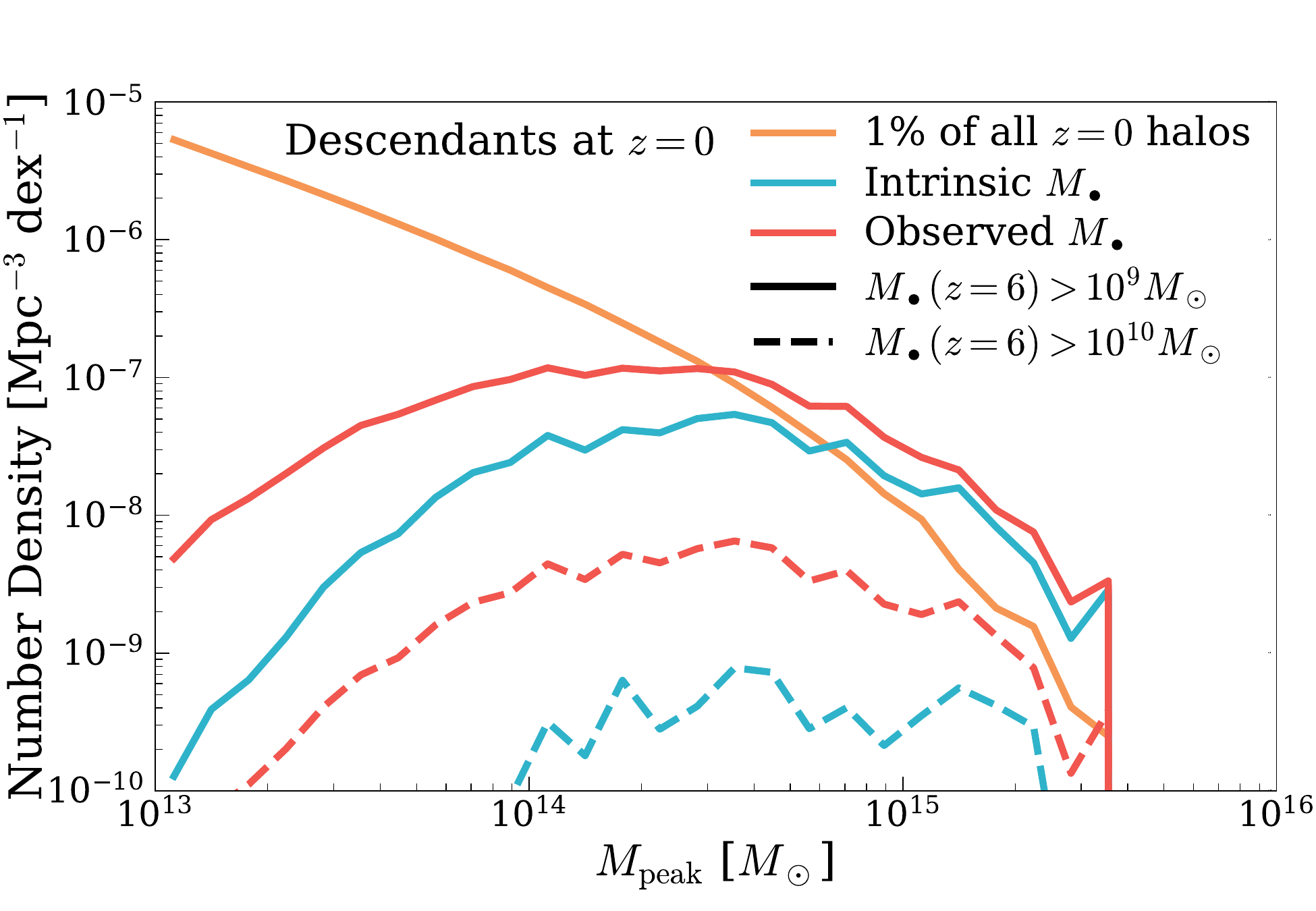}
}
\caption{\textbf{Top Panel:} The host halo mass functions of SMBHs with intrinsic (blue lines) and observed (red lines) $M_\bullet > 10^9 M_\odot$ (solid lines) and  $M_\bullet > 10^{10} M_\odot$ (dashed lines) at $z=6$, calculated by \textsc{Trinity}. The orange solid line is the $z=6$ halo mass function from the MDPL2 simulation, scaled down to 10\% of the original value for ease of comparison. \textbf{Bottom Panel:} The $z=0$ descendant halo mass functions of the $z=6$ haloes hosting SMBHs with intrinsic (blue lines) and red (red lines) $M_\bullet > 10^9 M_\odot$ (solid lines) and  $M_\bullet > 10^{10} M_\odot$ (dashed lines). They are calculated by following individual haloes in the MDPL2 simulation across time, and weighting each of them by the ratios between the host halo mass functions and the total halo mass functions in the top panel. The orange solid line is the $z=0$ halo mass function in the MDPL2 simulation, scaled down by a factor of 100. See \S\ref{ss:desc_big_mbh_at_z6}.}
\label{f:prog_desc_mbh9z6}
\end{figure}

We show the distribution of host halo masses for SMBHs with $M_\bullet > 10^9 M_\odot$ (solid curves) and $M_\bullet > 10^{10} M_\odot$ (dashed curves) at $z=6$, in the top panel of Fig.\ \ref{f:prog_desc_mbh9z6}. When using virial estimates to measure \mbh{}, there is random scatter in the measured \mbh{} around the intrinsic values. So, we also show the host halo mass functions for observed $M_\bullet > 10^9 M_\odot$ in blue and red curves, respectively. In this comparison, we assume a random scatter of $0.5$ dex in measured \mbh{}. This random scatter induces Eddington bias, i.e., more lower-mass haloes hosting \emph{intrinsically} smaller SMBHs have their \emph{observed} \mbh{} upscattered than the opposite. Thus, the host halo mass function (HMF) is both enhanced in normalization and shifted towards lower masses. The median host halo mass for SMBHs with intrinsic $M_\bullet > 10^9 M_\odot$ is $\sim 3\times 10^{12} M_\odot$, compared to $\sim 2\times 10^{12} M_\odot$ for SMBHs with observed $M_\bullet > 10^9 M_\odot$. For $M_\bullet > 10^{10} M_\odot$ SMBHs, their host haloes are typically $\sim 5\times 10^{12} M_\odot$ and $\sim 3\times 10^{12} M_\odot$, for intrinsic and observed $M_\bullet$, respectively. But overall, host haloes of $M_\bullet > 10^9(10^{10}) M_\odot$ SMBHs only make up $\sim 10\%(1\%)$ of all the haloes (orange solid curve, calculated from the MDPL2 simulation) at their respective median host halo masses, which correspond to $\sim 4.5-5.5 \sigma$ peaks in Gaussian random fields \citep{Eisenstein1998,Diemer2018}. In such haloes, gas reservoirs available for rapid SMBH growth could include cold flows (e.g., \citealt{Feng2014} and \citealt{Latif2022}), accretion triggered by protogalaxy mergers (e.g., \citealt{Li2007}), or a combination of the two (e.g., \citealt{Dubois2012}). Observationally, we expect to see overdensities of galaxies around the SMBHs in these massive haloes \citep{Wang2023}. In future work, we will create realistic mock galaxy+AGN catalogs to quantify the typical magnitude and stochastic variance of galaxy overdensities around $z\gtrsim 6$ quasars, and compare them with results from, e.g., the ASPIRE \citep{Wang2023} and EIGER \citep{Kashino2023} surveys.

Ever since the discovery of SMBHs with $M_\bullet > 10^9 M_\odot$, there have been studies investigating where their descendant SMBHs can be found in the local Universe \citep[e.g.,][]{Angulo2012,Fanidakis2013,Tenneti2018}.   To address this question with \textsc{Trinity}, we follow each $z=6$ halo in the MDPL2 simulation down to $z=0$, and examine their descendant halo masses. We weight each progenitor--descendant pair by the fraction of halos with the progenitor's mass at $z=6$ that would host a $>10^9/10^{10} M_\odot$ SMBH. As shown in the bottom panel of Fig.\ \ref{f:prog_desc_mbh9z6}, the descendant HMFs are much broader than the progenitor HMFs. This is due to the diversity in the assembly histories of individual haloes. At $z=0$, descendants with $M_\bullet (z=6) > 10^9 M_\odot$ typically have $M_\mathrm{peak} \sim 3-7 \times 10^{14} M_\odot$. However, these haloes only make up $\lesssim 1\%$ of the haloes at similar mass scales (the orange solid line). Combined with the small comoving volume in the local Universe, we would expect only $\sim 1$ such relic SMBH closer than $z=0.05$. For $M_\bullet (z=6) > 10^{10} M_\odot$, their host halo masses are similar to those with $M_\bullet (z=6) > 10^{9} M_\odot$, but they are much rarer (i.e., a factor of $\sim 30-100$). This makes it even harder to find the descendant haloes of these more massive high-redshift SMBHs in the local Universe.

\section{Discussion}
\label{s:discussion}

In \S\ref{ss:discussion_physics_bhsm}, we discuss the physical implications of the steeper intrinsic \bhsm{} relation between $7\lesssim z \lesssim 10$ from \textsc{Trinity}; In \S\ref{ss:discussion_z9_z11_agns}, we compare CEERS\_1019 and GN-z11 with more \textsc{Trinity} predictions at $6\lesssim z \lesssim 11$; In \S\ref{ss:compare_with_pacucci}, we compare the progenitor masses and growth rates of $\gtrsim 10^9 M_\odot$ SMBHs at $z\sim 9$ from \citet{Pacucci2022} with those from \textsc{Trinity}; \S\ref{ss:discussion_high_z_constraints} discusses the constraining power of $z=6-10$ quasar luminosity functions on the halo--galaxy--SMBH connection.

\subsection{Physical Implications of the \bhsm{} relation at $z\gtrsim 6$ from \textsc{Trinity}}
\label{ss:discussion_physics_bhsm}

According to \textsc{Trinity}, the intrinsic \bhsm{} relation between $7\lesssim z \lesssim 10$ is relatively static, with the largest difference being that the relation is mildy steeper at higher redshifts than at lower redshifts (see also \citealt{Volonteri2011}).  This means that SMBHs are less massive than in local galaxies at fixed stellar masses below $5\times 10^{10} M_\odot$ (\S\ref{ss:bhsm_bright_quasars}). Physically, this redshift evolution implies that SMBH growth in $M_*<10^{11} M_\odot$ galaxies is modestly slower than the host galaxy's growth. Several physical mechanisms could contribute to such limited growth: 1) strong supernova feedback in low-mass galaxies preventing efficient SMBH accretion (as in, e.g., IllustrisTNG, \citealt{Dubois2015,Bower2017,Pillepich2018}); 2) erratic motions of SMBHs due to shallow potential wells (as in, e.g., \citealt{Habouzit2017},  \citealt{Bellovary2019}, and \citealt{Dekel2019}, though c.f.\ \citealt{Choksi17}); and 3) Feedback from SMBHs themselves (e.g., as in \textsc{Romulus}, \citealt{Tremmel2017}). Limited SMBH growth at $z\gtrsim 7$ also implies that AGN are unlikely to dominate the cosmic reionization of intergalactic medium, as also suggested by, e.g., \citet{Finkelstein2019}.

\subsection{CEERS\_1019 and GN-z11 in the context of \textsc{Trinity}}
\label{ss:discussion_z9_z11_agns}

\subsubsection{SMBH mass and Eddington ratios}
\label{sss:z9_z11_mbh_eta}

\begin{figure}
\subfigure{
\includegraphics[width=0.48\textwidth]{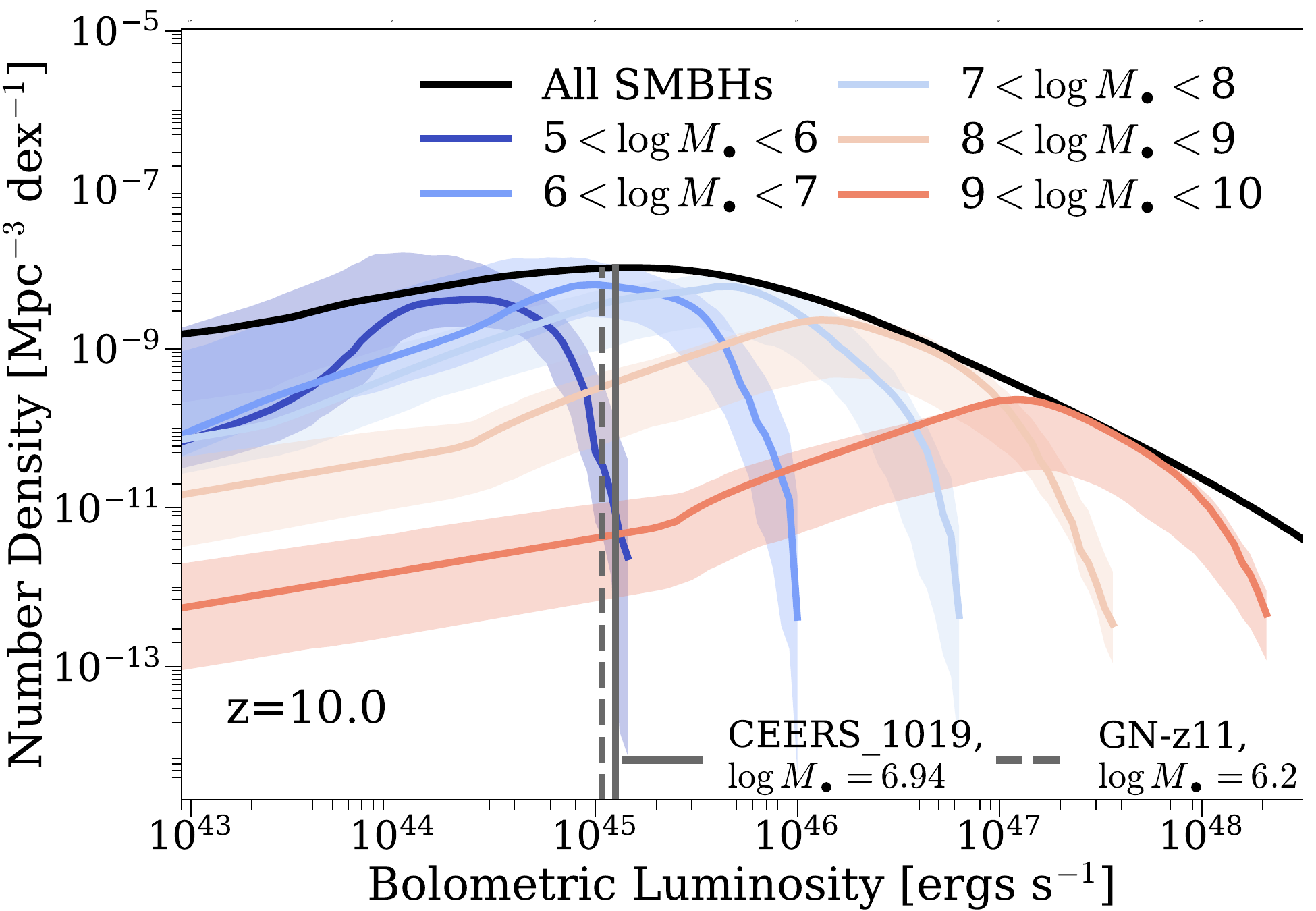}
}
\subfigure{
\includegraphics[width=0.48\textwidth]{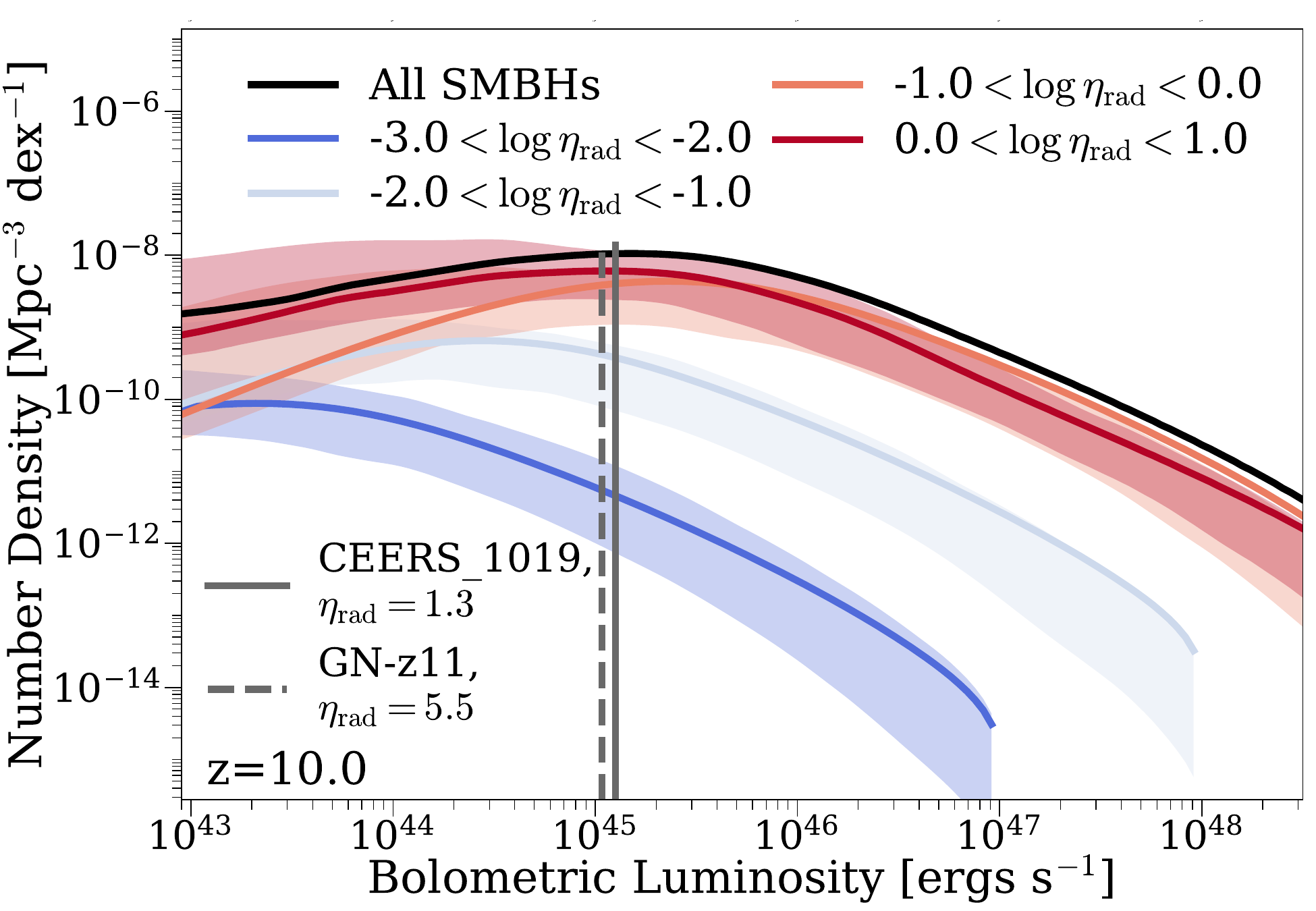}
}
\caption{\textbf{Top Panel:} The $z=10$ bolometric quasar luminosity function (QLF) decomposed into bins of different SMBH mass ($M_\bullet$). \textbf{Bottom Panel:} The $z=10$ bolometric quasar luminosity function (QLF) decomposed into bins of different SMBH Eddington ratio ($M_\bullet$). The solid and dashed vertical lines represent the bolometric luminosities of CEERS\_1019 and GN-z11, respectively. \shadedregions{} See \S\ref{sss:z9_z11_mbh_eta}.}
\label{f:qlf_mbh_eta_z10}
\end{figure}

Fig.\ \ref{f:qlf_mbh_eta_z10} shows the $z=10$ bolometric quasar luminosity function (QLF) decomposed into bins of SMBH mass ($M_\bullet$, top panel) and Eddington ratio ($\eta_\mathrm{rad}$), predicted by \textsc{Trinity}, which are relevant to help better understand the measured properties of CEERS\_1019 and GN-z11. At $z\sim 10$, the QLF peaks at around $10^{45}$ erg/s. In addition, SMBHs with $6 < \log M_\bullet < 7$ (top panel) and $0 < \log \eta_\mathrm{rad} < 1$ dominate the QLF at around $10^{45}$ erg/s. Given that both GN-z11 and CEERS1019 have bolometric luminosities of $\sim 10^{45}$ erg/s, their masses ($\log M_\bullet\sim 6.2$ and $6.94$) and Eddington ratios ($\eta_\mathrm{rad}\sim 5.5$ and $1.3$) are consistent with \textsc{Trinity}'s predictions (shown in the solid and dashed vertical lines). We also note that in \textsc{Trinity}, $z\sim10$ SMBHs grow at \textbf{slightly below} the Eddington rate, when averaged over both their active and dormant phases. Consequently, although GN-z11 is estimated to be accreting at a super-Eddington rate, such fast accretion phases are likely intermittent. Two objects are far from a statistically representative sample, so more detections and measurements of $z\gtrsim 9$ SMBHs are needed to further confirm/disprove our predictions on the dominant SMBH mass and Eddington ratios for these $z\gtrsim 8$ AGNs.

\subsubsection{Abundance of $M_\bullet > 10^{6}/10^{7} M_\odot$ SMBHs between $6<z<11$}
\label{sss:n_larson_maiolino_bh}

\begin{figure}
\includegraphics[width=0.48\textwidth]{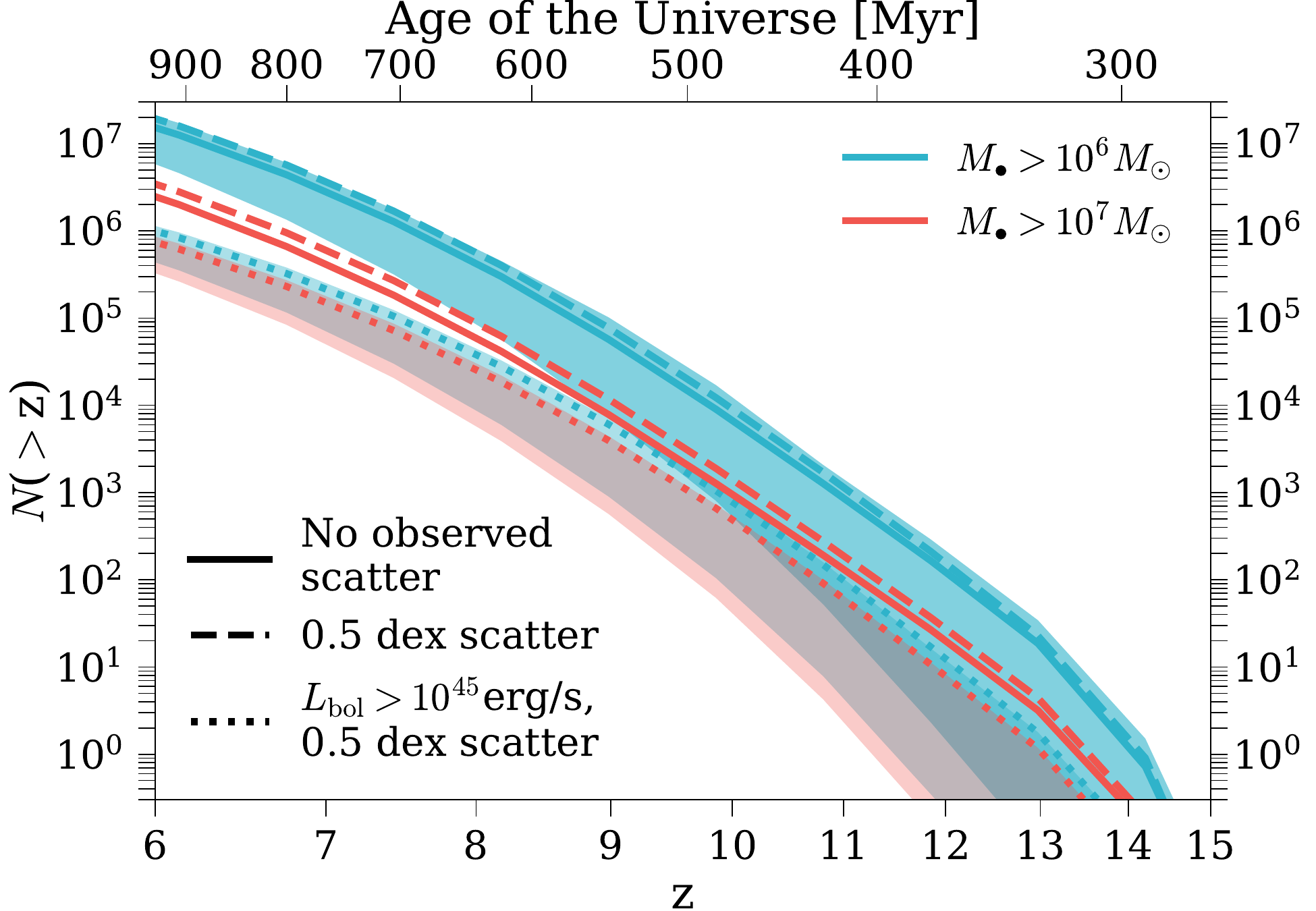}
\caption{The cumulative number of SMBHs with $M_\bullet>10^6 M_\odot$ (blue curves) and $M_\bullet>10^7 M_\odot$ (red curves). The line styles are the same as in Fig.\ \ref{f:huge_bh}, but we adopt a constant luminosity limit of $10^{45}$ erg/s in this figure. The darker blue shaded region is the 68$^{\mathrm{th}}$ percentile range (from MCMC) for $M_\bullet>10^6 M_\odot$ with no scatter in $M_{\bullet,\mathrm{obs}}$. The \emph{relative} uncertainty in the number of $M_\bullet>10^7 M_\odot$ SMBHs (red solid curve) is very similar to $M_\bullet>10^6 M_\odot$ SMBHs, so we do not show its confidence intervals for visual clarity. and the lighter blue(red) shaded region is the the 68$^{\mathrm{th}}$ percentile range for $M_\bullet>10^6 M_\odot$($10^7 M_\odot$) above the bolometric luminosity of $10^{45}$ erg/s, with $M_{\bullet,\mathrm{obs}}=0.5$ dex. See \S\ref{sss:n_larson_maiolino_bh}.}
\label{f:n_larson_maiolino_bh}
\end{figure}

Fig.\ \ref{f:n_larson_maiolino_bh} shows the cumulative number of SMBHs above $10^6 M_\odot$ (blue curves) and $10^7 M_\odot$ (red curves) from $z=6-11$. The solid curves represent the number of all (i.e., active+dormant) SMBHs with \emph{intrinsic} $M_\bullet$ above these mass limits, whereas the dashed curves denote the number of all SMBHs with \emph{observed} $M_\bullet$ above the limits, assuming a random scatter of 0.5 dex in observed $M_\bullet$. The dotted curves are similar to the dashed curves, but for SMBHs above the bolometric luminosity limit of $10^{45}$ erg/s. This limit has been chosen to match the luminosities of CEERS1019 and GN-z11. Similar to Fig.\ \ref{f:huge_bh}, these curves represent the numbers of SMBHs we expect to find \emph{over the full sky}. Compared to more massive SMBHs in Fig.\ \ref{f:huge_bh}, the scatter in observed $M_\bullet$ causes a smaller boost in the number of SMBHs, i.e., a factor of $\sim 1.5$. This is because the total BHMF is flatter at the less massive end, which reduces the magnitude of Eddington bias. According to \textsc{Trinity}, we expect to find $\sim 6000$ and $\sim 1500$ SMBHs with observed $M_\bullet>10^6 M_\odot$ and $10^7 M_\odot$ at $z>10$, respectively. However, only part of the underlying SMBH populations are bright enough to be detected. Given the nearly mass-independent Eddington ratio distributions from \textsc{Trinity}, fewer less massive SMBHs will survive the same luminosity limit. Consequently, when the luminosity limit is applied, the numbers of SMBH with $M_\bullet>10^6 M_\odot$ and $10^7 M_\odot$ at $z>10$ decrease to $\sim 600$ and $\sim 400$ over the whole sky, respectively.

\subsubsection{Host halos of CEERS1019 and GN-z11}
\label{sss:host_halo_larson_maiolino}

\begin{figure}
\subfigure{
\includegraphics[width=0.48\textwidth]{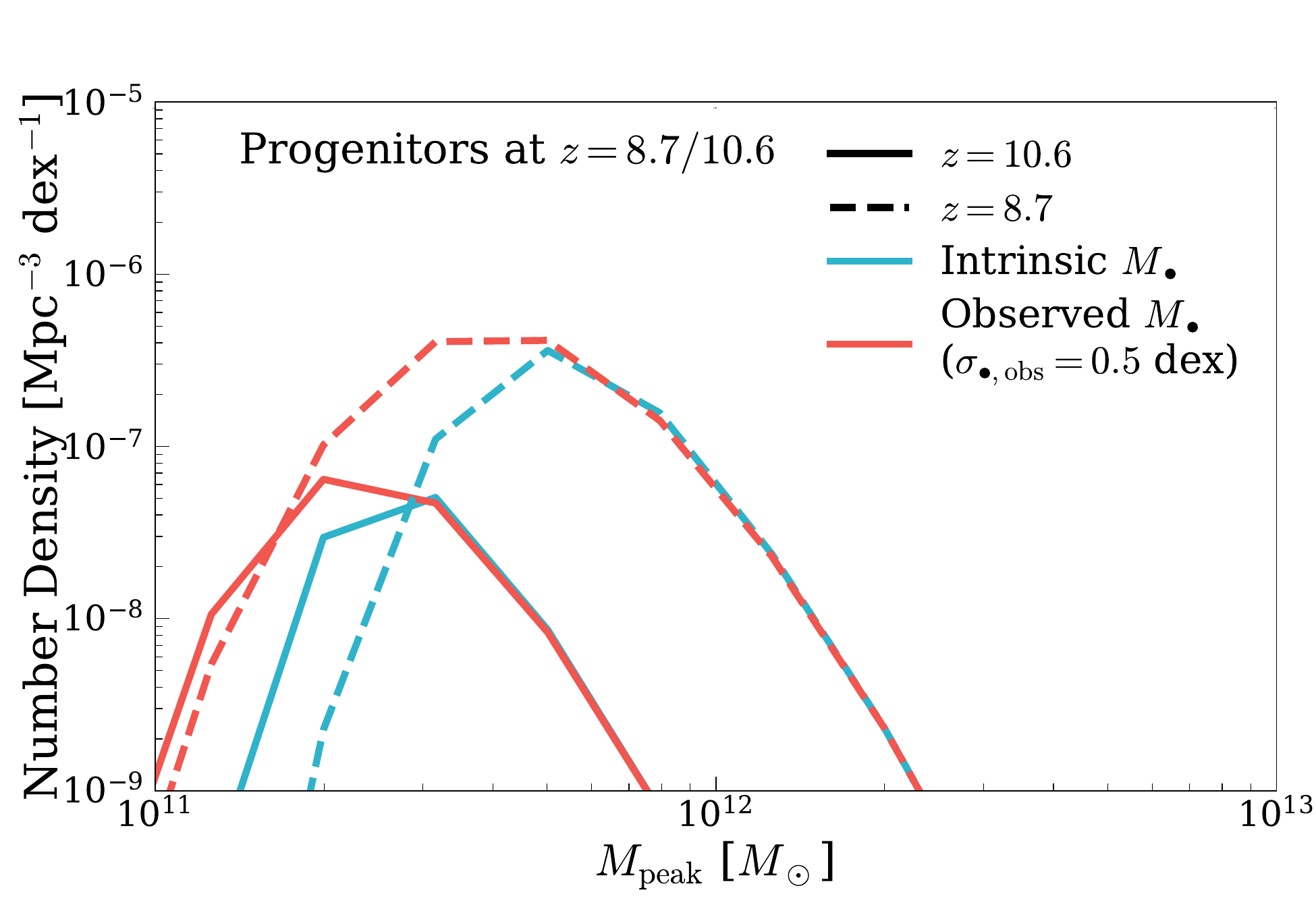}
}
\subfigure{
\includegraphics[width=0.48\textwidth]{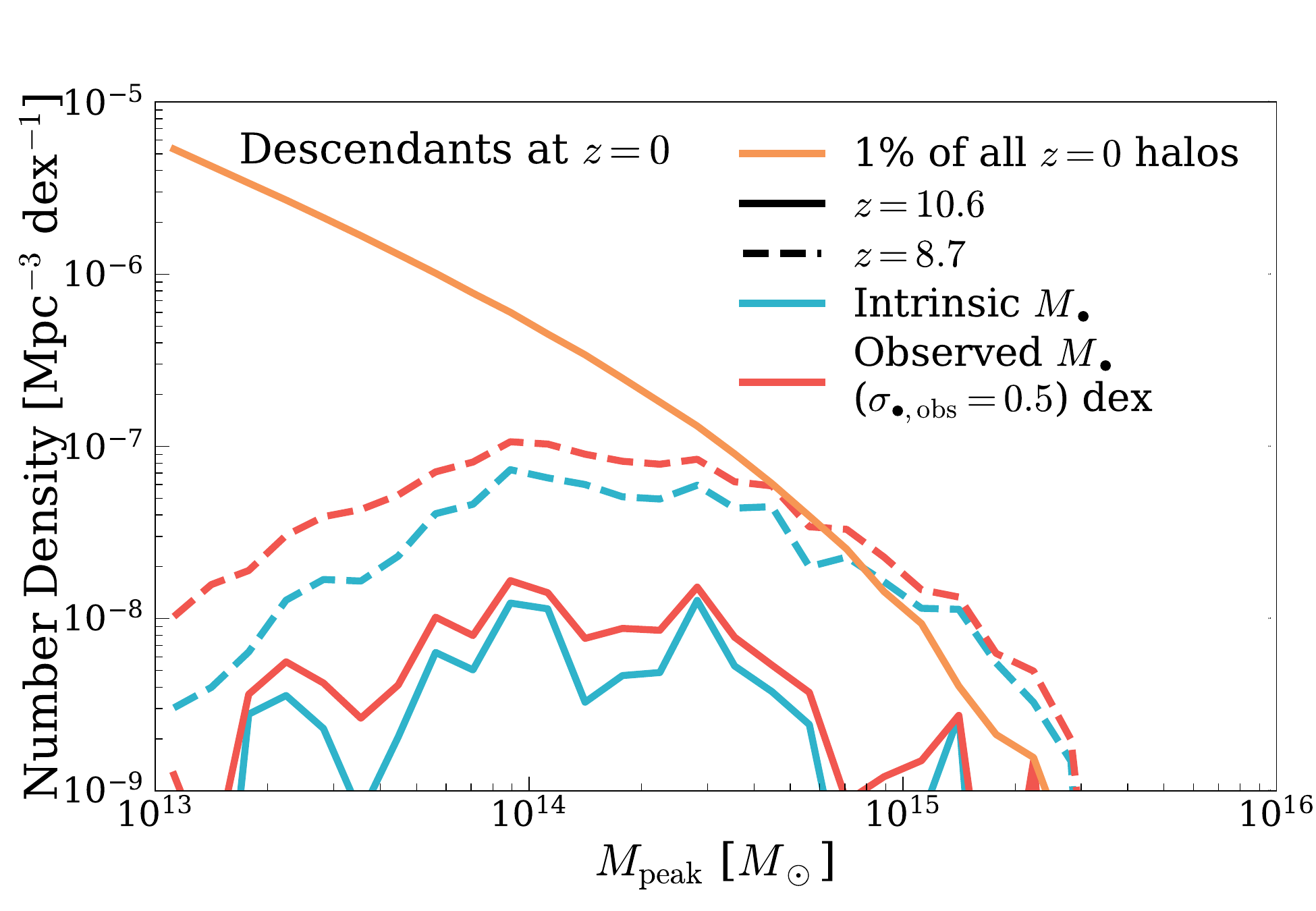}
}
\caption{\textbf{Top Panel:} The host halo mass functions of SMBHs with intrinsic (blue curves) and observed (red curves) $M_\bullet>10^6 M_\odot$ at $z=10.6$ (solid curves) and $M_\bullet>10^7 M_\odot$ at $z=8.7$ (dashed curves). \textbf{Bottom Panel:} The $z=0$ descendant halo mass functions of halos hosting ntrinsic (blue curves) and observed (red curves) $M_\bullet>10^6 M_\odot$ at $z=10.6$ (solid curves) and $M_\bullet>10^7 M_\odot$ at $z=8.7$ (dashed curves). The orange solid curve is the total host halo mass functions $z=0$, scaled down by a factor of 100. See \S\ref{sss:host_halo_larson_maiolino}.}
\label{f:host_halo_larson_maiolino}
\end{figure}

In the top panel of Fig.\ \ref{f:host_halo_larson_maiolino}, we show the host halo mass functions of SMBHs with intrinsic (blue curves) and observed (red curves) $M_\bullet>10^6 M_\odot$ at $z=10.6$ (solid curves) and $M_\bullet>10^7 M_\odot$ at $z=8.7$ (dashed curves). These redshifts and mass limits are chosen to roughly match the observed properties of CEERS\_1019 and GN-z11. Similar to the host halo mass functions of $M_\bullet>10^9 M_\odot$ at $z=6$ (\S\ref{ss:desc_big_mbh_at_z6}, Fig.\ \ref{f:prog_desc_mbh9z6}), the difference between the red and blue curves comes from the Eddington bias that upscatters undermassive halos hosting intrinsically smaller SMBHs. GN-z11 is likely to be hosted in a $(2-3) \times 10^{11} M_\odot$ halo, whereas CEERS1019's typical host halo mass is $(3-6)\times 10^{11} M_\odot$. These host halo masses correspond to $\gtrsim 7\sigma$ peaks in Gaussian random fields at their respective redshifts, making them candidates for protocluster cores. This means that the typical host haloes for CEERS\_1019 and GN-z11 are rarer than the luminous quasars at $z\sim 6$ at their respective redshifts. We note that our estimated host halo mass for GN-z11 is $\sim 1$ dex higher than the one by \citet{Scholtz2023}. Scholtz et al.\ made this estimate based on the stellar mass measurement by \citet{Tacchella2023} and the stellar mass--halo mass relation from \citet{Behroozi2019}. However, \textsc{Trinity} predicts that the AGN duty cycle in low mass haloes are much lower. In other words, low mass haloes are much less likely to host AGNs like GN-z11. These predictions are driven by extrapolating the QPDFs from \citet{Aird2018} to high redshifts. Therefore, \textsc{Trinity} finds more massive host haloes for GN-z11, despite the very similar galaxy mass--halo mass relations at $z\gtrsim 10$.

The bottom panel of Fig.\ \ref{f:host_halo_larson_maiolino} shows the $z=0$ descendant halo mass functions of haloes hosting $M_\bullet>10^6 M_\odot$ at $z=10.6$ and $M_\bullet>10^7 M_\odot$ at $z=8.7$. These descendant halo mass functions are calculated using the method introduced in \S\ref{ss:desc_big_mbh_at_z6}. The typical descendant halo masses for GN-z11- and CEERS\_1019-like objects are $\sim (1-3)\times 10^{14} M_\odot$, although the distribution is broad, due to the diversity in individual halo assembly histories. Similar to the $z=0$ descendants of $z\sim 6$ massive black holes, the descendants of GN-z11- and CEERS\_1019-like objects make up only $\lesssim 1\%$ of all the haloes at similar mass scales. This makes them too rare to find in the local Universe. Compared to the descendant halo mass of GN-z11 obtained by \citet{Scholtz2023} based on the Millenium Simulation \citep{Springel2005,Chiang2013}, our descendant mass scale is lower by $\sim 1$ dex. This is because \citet{Scholtz2023} chose to look at the average $z=10.6$ \emph{progenitor} halo masses of $z=0$ haloes above $10^{15} M_\odot$, and found that the average progenitor halo mass coincides with the halo mass estimate of GNz-11. Due to the scatter in individual halo assembly histories, this is different from looking at the $z=0$ \emph{descendant} halo masses of halo populations with certain masses at higher redshifts, and will lead to a larger $z=0$ descendant halo mass. In this work, we chose the opposite method because it is more representative of the expected descendant masses; the \citet{Scholtz2023} method is more appropriate for finding $z=0$ objects whose typical progenitors look like GNz-11.

\subsubsection{Potential descendants and progenitors of CEERS1019 and GN-z11}
\label{sss:growth_track_larson_maiolino}

\begin{figure}
\subfigure{
\includegraphics[width=0.48\textwidth]{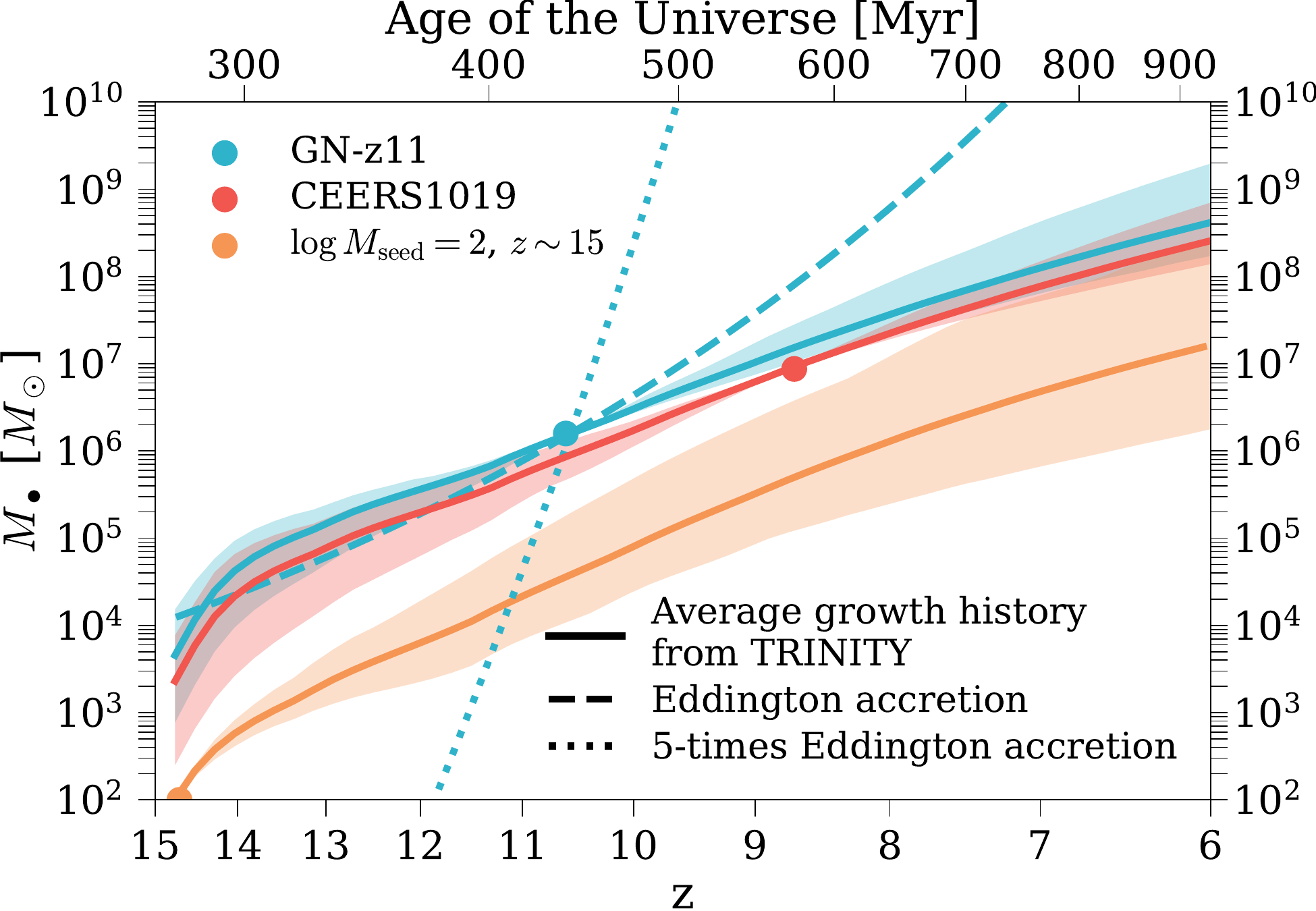}
}
\caption{The \emph{average} mass growth curves of GN-z11 (the blue solid curve and circle), CEERS1019 (the red curve and circle), and a seed with $M_\mathrm{seed}=100 M_\odot$ at $z\sim 15$ (the orange curve and circle), predicted by \textsc{Trinity}. The blue dashed curve shows the mass growth history of GN-z11 assuming Eddington-limited accretion. The shaded regions reflect the statistical uncertainties in the MCMC process in \textsc{Trinity}, given the existing data constraints. See \S\ref{sss:growth_track_larson_maiolino}.}
\label{f:growth_track_larson_maiolino}
\end{figure}

To further understand the potential progenitors and descendants of GN-z11 and CEERS\_1019, we calculated the \emph{average}\footnote{I.e., averaged over all active and dormant SMBHs of the same mass at the same time} Eddington ratio as a function of $M_\bullet$ and time, and calculated the average mass growth curves (shown in Fig.\ \ref{f:growth_track_larson_maiolino} for: 1) GN-z11 (blue solid curve); 2) CEERS\_1019 (red solid curve); and 3) a seed of $100 M_\odot$ at $z\sim 15$ (orange solid curve). The shaded regions represent the statistical uncertainties from the MCMC, which reflects the uncertainty in the input observational constraints (see \S\ref{s:method}). In other words, they quantify how uncertain the \emph{average} mass growth curves are given all the existing data constraints, but \emph{does not} show the scatter in the growth histories among \emph{individual} SMBHs. The blue dashed(dotted) curve shows the SMBH mass growth with a constant Eddington ratio of 1(5). 

The high Eddington ratios at $z\gtrsim 6$ mean that SMBHs grow exponentially, although the $e$-folding time scale increases due to the decline in Eddington ratio with time. Consequently, the uncertainty in $M_\bullet$ increases as we move away from the starting redshift. This effect is the most significant for the $100 M_\odot$ seed, whose growth was traced for more than 600 Myrs in one single direction. As a comparison, the average growth histories of GN-z11 and CEERS\_1019 are only traced for at most $\sim 450$ Myrs in each direction. At $z\sim10.5$ and $z\sim 8.7$, the uncertainty in the $100 M_\odot$ seed's descendant mass does not cover either GN-z11 or CEERS\_1019. By $z=6$, the uncertainty in its descendant mass is $\sim 2$ dex. This means that with current observational constraints on \textsc{Trinity}, the potential progenitors of GN-z11, CEERS\_1019 are unlikely to be as small as $100 M_\odot$ at $z\sim 15$.

\emph{On average}, SMBHs grow at above(below) the Eddington rate above(below) $z\sim 13$. Thus, even though GN-z11 is accreting at five times the Eddington rate at the time of observation, such fast accretion episodes are likely intermittent (also see \S\ref{sss:z9_z11_mbh_eta}). This difference between the average and the instantaneous Eddington ratio is vital in estimating SMBH growth histories, given the exponential nature of early SMBH growth. As shown in Fig.\ \ref{f:growth_track_larson_maiolino}, a constant Eddington ratio of 5 enables the fast mass buildup of $\sim 10^{10} M_\odot$ within $\sim 120$ Myrs. Conversely, it would take $\sim 500$ Myrs to achieve the same amount of mass growth with the Eddington rate. The drastically different SMBH growth histories also point to different SMBH progenitor masses, e.g., the constant super-Eddington growth implies the possibility of creating a GN-z11-like SMBH from with $100 M_\odot$ progenitor at $z\sim 12$, which is strongly disfavored by the constant Eddington-limited scenario. In light of this, it is critical to use more accurate estimates of average Eddington ratios, rather than instantaneous values, when inferring potential SMBH progenitor masses. Because it is not possible to measure the average historical Eddington ratios for individual SMBH, it is essential to have models like \textsc{Trinity} that infer average Eddington ratios for high-redshift SMBHs, based on as many data constraints as possible.

We then discuss the potential seeding mechanisms for early SMBHs like GN-z11 and CEERS\_1019. As can be seen from Fig.\ \ref{f:growth_track_larson_maiolino}, the typical $z\sim 15$ progenitor masses of GN-z11 and CEERS\_1019 are $\sim 10^3-10^4 M_\odot$, although CEERS\_1019 could be descendants of seeds as small as $\sim 300 M_\odot$ at $z\sim 15$. Such a progenitor mass range is consistent with the scenario where the seeds are created via runaway mergers in nuclear star clusters (see, e.g., \citealt{Devecchi2012,Lupi2014,Schleicher2022}). In addition, when SMBH seeds form in star clusters, their dynamics are determined by the cluster mass rather than the seed mass, which makes them less susceptible to erratic motions than PopIII remnants. Compared to PopIII remnant seeds, they are thus also more likely to stay in high gas density regions and experience consistent growth as predicted by \textsc{Trinity}. Consequently, we believe that if the SMBHs are seeded at $z\sim 15$, the star cluster scenario is more plausible for objects like GN-z11 and CEERS\_1019, compared to the PopIII remnant scenario. We also note that since the $68^{\mathrm{th}}$ percentile range of GN-z11's and CEERS\_1019's progenitor masses cover $\sim 10^4 M_\odot$ at $z\sim 11-14$, we \emph{cannot} rule out the supermassive star scenario that would produce such heavy seed SMBHs (see, e.g., \citealt{Inayoshi2020} and the references therein).

\subsubsection{AGN fraction among $6<z<11$ galaxies}
\label{sss:highz_agn_fraction}

\begin{figure}
\includegraphics[width=0.48\textwidth]{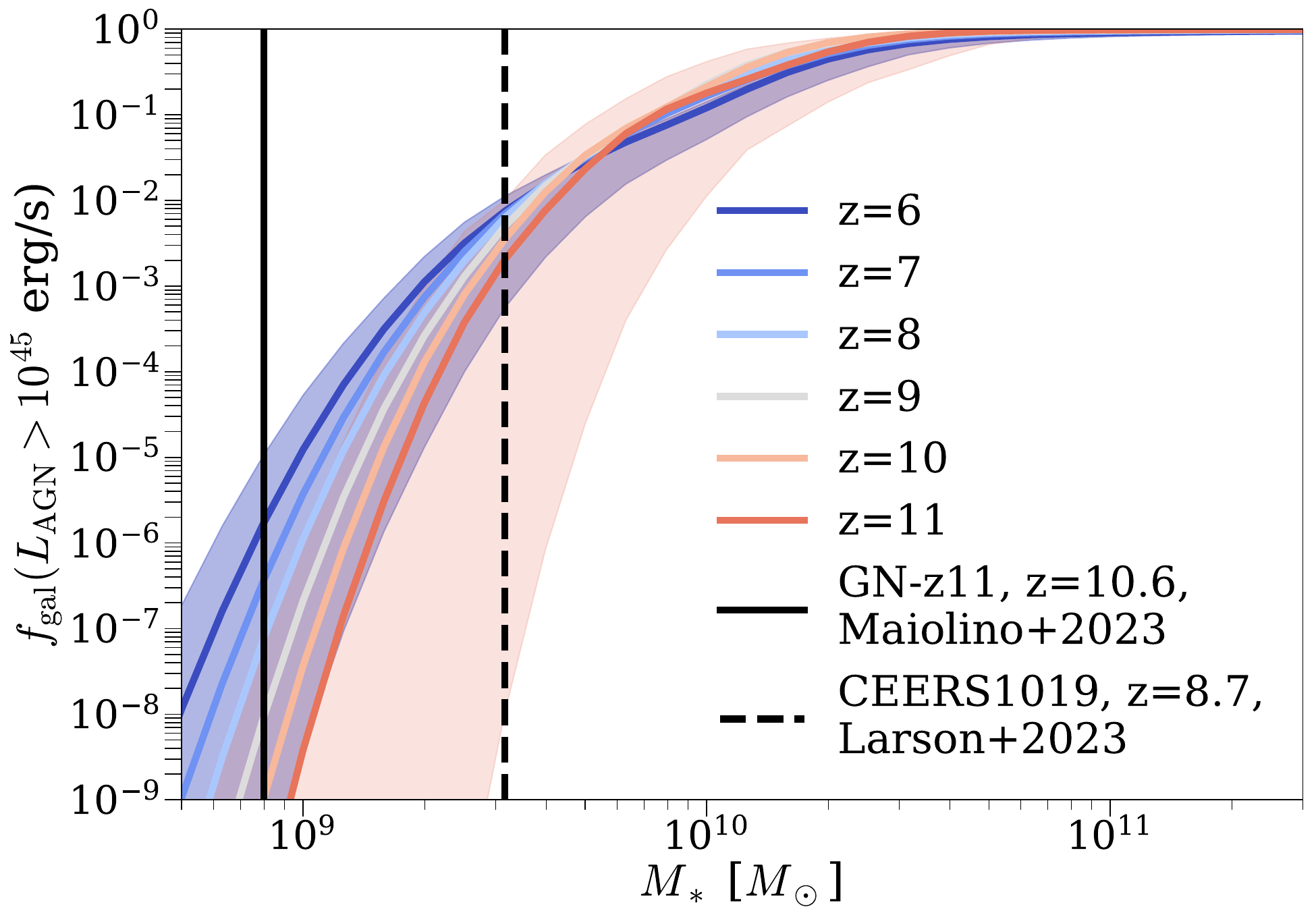}
\caption{The fraction of galaxies that host AGNs above the bolometric luminosity of $10^{45}$ erg/s as a function of galaxy mass ($M_*$) and redshift. The solid and dashed vertical lines denote the estimated galaxy masses for \citet{Maiolino2023} and \citet{Larson2023}, respectively. For visual clarity, we only show the 68$^{\mathrm{th}}$ percentile ranges (from MCMC) for the $z=6$  blue shaded region) and the $z=11$ (orange shaded region) cases. See \S\ref{sss:highz_agn_fraction}.}
\label{f:highz_agn_frac}
\end{figure}

Fig.\ \ref{f:highz_agn_frac} shows the fraction of galaxies hosting AGNs with bolometric luminosities above $10^{45}$ erg/s, $f_\mathrm{gal}$, as a function of galaxy mass and redshift. These fractions include both obscured (including Compton-thick obscured) and non-obscured AGNs. The stochastic uncertainties in $f_\mathrm{gal}$ at $z=6$($z=11$) are shown in blue(orange) shaded regions. We also show the galaxy masses of GN-z11 and CEERS\_1019 in red solid and dashed lines, respetively.

At $z\gtrsim 6$,  $f_\mathrm{gal}$ is a strong function of galaxy mass, increasing from $\lesssim 10^{-6}$ at $M_*\sim 10^9 M_\odot$ to $\sim 100\%$ at $M_*\sim 3\times 10^{10} M_\odot$. This is an extrapolation based on quasar probability distribution functions (QPDFs) from \citet{Aird2018}, which requires AGN duty cycle to be low in less massive galaxies. In addition, the AGNs in Fig.\ \ref{f:highz_agn_frac} are selected with a fixed luminosity. The fact that lower-mass galaxies host smaller SMBHs also naturally decreases the incidence of such AGNs at the low-mass end. The high incidence of AGN in massive galaxies suggests that it is essential to properly account for any AGN contribution when trying to characterize high-redshift massive galaxies \citep{Volonteri2017}. On the flip side, it also means that detections of high-redshift massive galaxies that have not included AGN contributions likely need to be revisited (e.g., \citealt{Labbe2023}). In the low-mass regime, it is \emph{statistically} more acceptable to ignore \emph{potential} AGN contributions when characterizing galaxies. However, it is still essential to account for AGN contributions when characterizing host galaxies of \emph{confirmed} AGNs like GN-z11 and CEERS\_1019. It also implies that GN-z11 and CEERS\_1019 are rare cases where low-mass galaxies do host bright AGNs, which might be in tension with observations given the survey areas of JADES and CEERS (see also \citealt{Maiolino2023b} for an estimate of AGN fraction). As mentioned in \S\ref{sss:n_larson_maiolino_bh}, in addition to a better understanding of galaxy mass measurements and various selection effects, non-detection over a much bigger survey area or a larger sample of low-mass galaxies with similar AGNs is needed to confirm/disprove this prediction from \textsc{Trinity}, as well as provide valuable constraints on $f_\mathrm{gal}$ at high redshifts.

\subsubsection{Using CEERS\_1019 and GN-z11 as constraints on \textsc{Trinity}}
We conclude \S\ref{ss:discussion_z9_z11_agns} by quantifying the constraint that CEERS\_1019 and GN-z11 can impose on the current \textsc{Trinity} model when added to the current data compilation. Given the uncertainties in their host galaxy masses, we chose not to compare them with the number density of AGNs hosted by similar-mass galaxies, as in Fig.\ \ref{f:highz_agn_frac}. Instead, we converted the detection of CEERS\_1019 and GN-z11 into the cumulative number density of AGNs above $L_\mathrm{bol}=10^{45}$ erg/s between $z=8-12$. To calculate the comoving volume, we combined the survey areas of CEERS, GOODS-S, and GOODS-N fields, which amount to 260 arcmin$^2$. The fiducial \textsc{Trinity} model predicts $\sim 0.15$ AGNs above $L_\mathrm{bol}=10^{45}$ erg/s between $z=8-12$ in these fields. Using the two detected AGNs as a lower limit and assuming Poisson statistics, this corresponds to a $\chi_{z>9}^2\sim 9.2$. We added this additional number density constraint to \textsc{Trinity}, and reran the MCMC. We found that $\chi_{z>9}^2$ is too small to significantly change the posterior distribution of model parameters, compared to the total $\chi_\mathrm{tot}^2\sim 760$ from other data. In other words, this means that the constraint imposed by the AGN number density from CEERS\_1019 and GN-z11 is not strong enough to change the empirical galaxy--SMBH connection that is already constrained by all the other existing galaxy and SMBH data. This further confirms that we need either non-detections of such high-redshift AGNs over a larger volume and better host galaxy measurements to confirm \textsc{Trinity}'s predictions, or more detections per unit volume to provide strong enough constraints on \textsc{Trinity}.

If future observations (with selection effects corrected) confirmed that the AGN fraction is indeed much higher for  $M_* \lesssim 10^9 M_\odot$ galaxies than predicted by \textsc{Trinity} (Fig.\ \ref{f:highz_agn_frac}), then \textsc{Trinity} would likely have underestimated the following SMBH properties: 1) the typical SMBH masses; 2) typical AGN Eddington ratios; and/or 3) AGN duty cycles in these low-mass galaxies. Physically, this implies that high-redshift SMBHs in low-mass galaxies likely grow faster and more consistently than their low-redshift counterparts, which is not captured by extrapolating low-redshift data with the current parameterization. One possible solution to this tension is to introduce new model parameters to describe the high-redshift behavior of these SMBH properties that are not captured by the current extrapolation. We will defer this exploration until we have a sufficient number of $z\gtrsim 8$ AGNs with robust measurements of host galaxy properties.

\subsection{Constraints on the progenitor masses and growth rates of $\gtrsim 10^9 M_\odot$ SMBHs at $z\sim 9$}
\label{ss:compare_with_pacucci}

In this section, we compare the potential progenitor masses and growth rates of $\gtrsim 10^9 M_\odot$ SMBHs from \citet{Pacucci2022} and \textsc{Trinity}. Since the methodologies of \textsc{Trinity} and the Monte Carlo simulation by Pacucci \& Loeb are very different, we will only make qualitative statements on any differences, to avoid over-interpretation. \citet{Pacucci2022} argued that the detection of $\gtrsim 10^9 M_\odot$ SMBHs at $z\gtrsim9$ would strongly disfavour the PopIII remnant seeding mechanism, due to the insufficient time to grow substantially within the Eddington limit. Based on their Monte Carlo simulation, they conclude that the mean seed mass, $M_\mathrm{seed}$, for $\geq 10^9 M_\odot$ SMBHs at $z\gtrsim9$ is $\log M_\mathrm{seed} = 5.34$ at $z\sim 25$. Since the \textsc{Trinity} model starts at $z\sim 15$, it is impossible to extrapolate towards higher redshifts with \textsc{Trinity}. We thus opt to redo the Monte Carlo simulation as described in \citet{Pacucci2022}, and compare the $z\sim15$ progenitor masses of $\gtrsim 10^9 M_\odot$ SMBHs at $z\gtrsim9$ with \textsc{Trinity}. 

We find after this exercise that the $z\sim15$ progenitor masses peak around $\log M_\mathrm{\bullet,z\sim 15}\sim 7$. To calculate the $z\sim15$ progenitor masses from \textsc{Trinity}, we make use of the predicted mass-independence of the Eddington ratio at high redshifts. Specifically, the $z\sim 9$ descendant of GN-z11 is predicted to have $\log M_\bullet\sim 7.5$, according to Fig.\ \ref{f:growth_track_larson_maiolino}. Its $z\sim 15$ progenitor has a mass of $\log M_\bullet\sim 3$. Given that Eddington ratio is nearly mass-independent, a $z\sim 15$ SMBH must be at least $\log M_\bullet\sim 3 + (9 - 7.5) = 4.5$ to reach $10^9 M_\odot$ by $z\sim 9$. This is roughly 2.5 dex lower than the typical $z\sim 15$ progenitor masses from \citet{Pacucci2022}. Several factors could lead to this difference: 1) \citet{Pacucci2022} adopted a \emph{flat} prior in $\log M_\mathrm{seed}$, which is very different from the high-redshift BHMFs inferred by \textsc{Trinity} (see Fig.\ \ref{f:bhmf}). Consequently, the massive seeds will be given much larger weights in \citet{Pacucci2022} than in \textsc{Trinity}; 2) In \citet{Pacucci2022}, SMBHs are not allowed to accrete at super-Eddington rates at any time. On the contrary, SMBHs are predicted to accrete at super-Eddington rates at $z\gtrsim 10.5$ in \textsc{Trinity}, even when averaged over active and dormant phases. This allows smaller SMBHs to grow into $\gtrsim 10^9 M_\odot$ SMBHs by $z\gtrsim9$ in \textsc{Trinity}.

With constant or slowly changing Eddington ratios, SMBH experience exponential growth:

\begin{equation}
    M_\bullet (t) = M_\mathrm{seed}\times \exp(\mathcal{D} \eta \frac{1 - \epsilon}{\epsilon} \frac{\Delta t}{t_\mathrm{Edd}})\ ,
\end{equation}
where $\mathcal{D}$ is the AGN duty cycle, i.e., the fraction of time during which SMBHs are accreting at an Eddington ratio $\eta$. $\epsilon$ is the AGN radiative efficiency, $\Delta t$ is the time interval between SMBH seeding and the time of observation $t$, and $t_\mathrm{Edd}\equiv 450$ Myr. Since it is the product of $\mathcal{D}$, $\eta$, and $(1 - \epsilon)/\epsilon$ that determines the growth rate, we define:
\begin{equation}
    \lambda \equiv \mathcal{D} \eta \frac{1 - \epsilon}{\epsilon}\ ,
\end{equation}
so that 
\begin{equation}
    M_\bullet (t) = M_\mathrm{seed}\times \exp(\lambda \frac{\Delta t} {t_\mathrm{Edd}})\ ,
\end{equation}
and compare the average $\lambda$ values from \textsc{Trinity} with the peak value of the $\lambda$ distribution from the Monte Carlo simulation by \citet{Pacucci2022}. Among $\gtrsim 10^9 M_\odot$ SMBHs by $z\sim 9$, the $\lambda$ distribution from \citet{Pacucci2022} peaks at $\lambda\sim 8.5$. In \textsc{Trinity}, the average $\lambda$ decreases from $\sim 60$ at $z\sim 15$ to $\sim 8.8$ by $z\sim 9$. In other words, SMBHs in \textsc{Trinity} grow significantly faster than simulated by \citet{Pacucci2022}. Given the similar AGN efficiency and duty cycle values, this difference is mostly caused by allowing/disallowing super-Eddington accretion in \textsc{Trinity}/\citet{Pacucci2022}. As mentioned in the last paragraph, this difference in growth rate directly affects the plausible progenitor masses for $\gtrsim 10^9 M_\odot$ SMBHs by $z\sim 9$ and could give different implications on SMBH seeding mechanisms. Our comparison here demonstrates again the importance of characterizing early SMBH growth rates in determining their seeding scenarios.

\subsection{Constraining power of $z\gtrsim 6$ quasar luminosity functions}
\label{ss:discussion_high_z_constraints}

As mentioned in \S\ref{ss:justification}, there is currently no $z>6.5$ SMBH data in the observational constraints for \textsc{Trinity}. Therefore, our predictions at $z\gtrsim 7$ are effectively extrapolations based on SMBH data at $z\leq 6.5$, in addition to galaxy data between $0<z<10$. Consequently, we see increasing statistical uncertainties towards higher redshifts in many predictions (e.g., Figs.\ \ref{f:huge_bh} and \ref{f:bhmf}). However, new instruments like JWST, \textit{Roman Space Telescope}, and \textit{Euclid} will provide an unprecedented amount of $z>6.5$ SMBH observations. Even before these new instruments, large ground-based telescopes already give constraints on QLFs up to $z\sim 7$ (e.g., \citealt{Matsuoka2023}). These high-redshift data will be vital for understanding SMBH formation and evolution in the early Universe. In this section, we demonstrate this by quantifying the constraining power of $z>6.5$ quasar luminosity functions (QLFs) on the halo--galaxy--SMBH connection in \textsc{Trinity}.

\begin{figure}
\subfigure{
\includegraphics[width=0.48\textwidth]{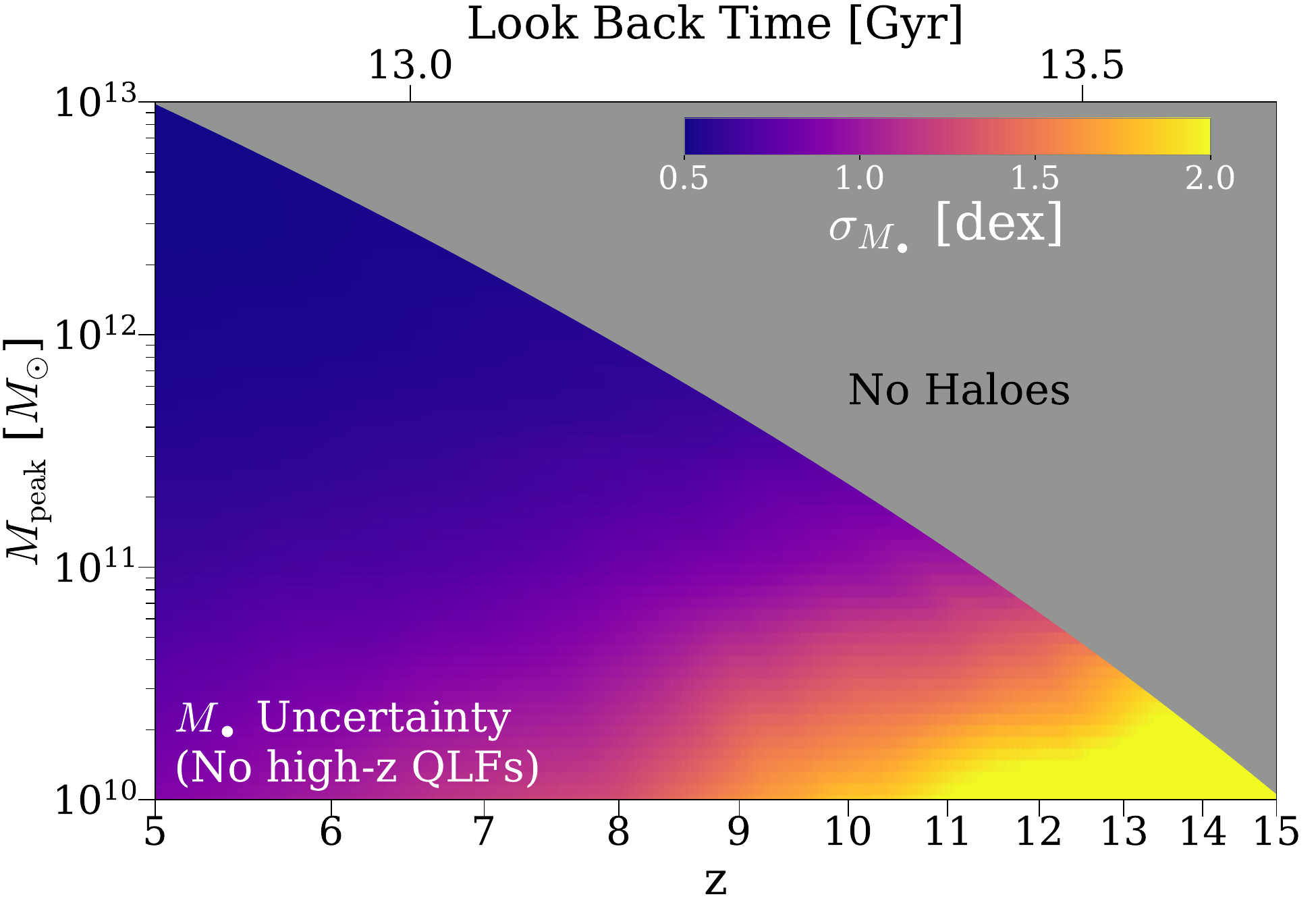}
}
\subfigure{
\includegraphics[width=0.48\textwidth]{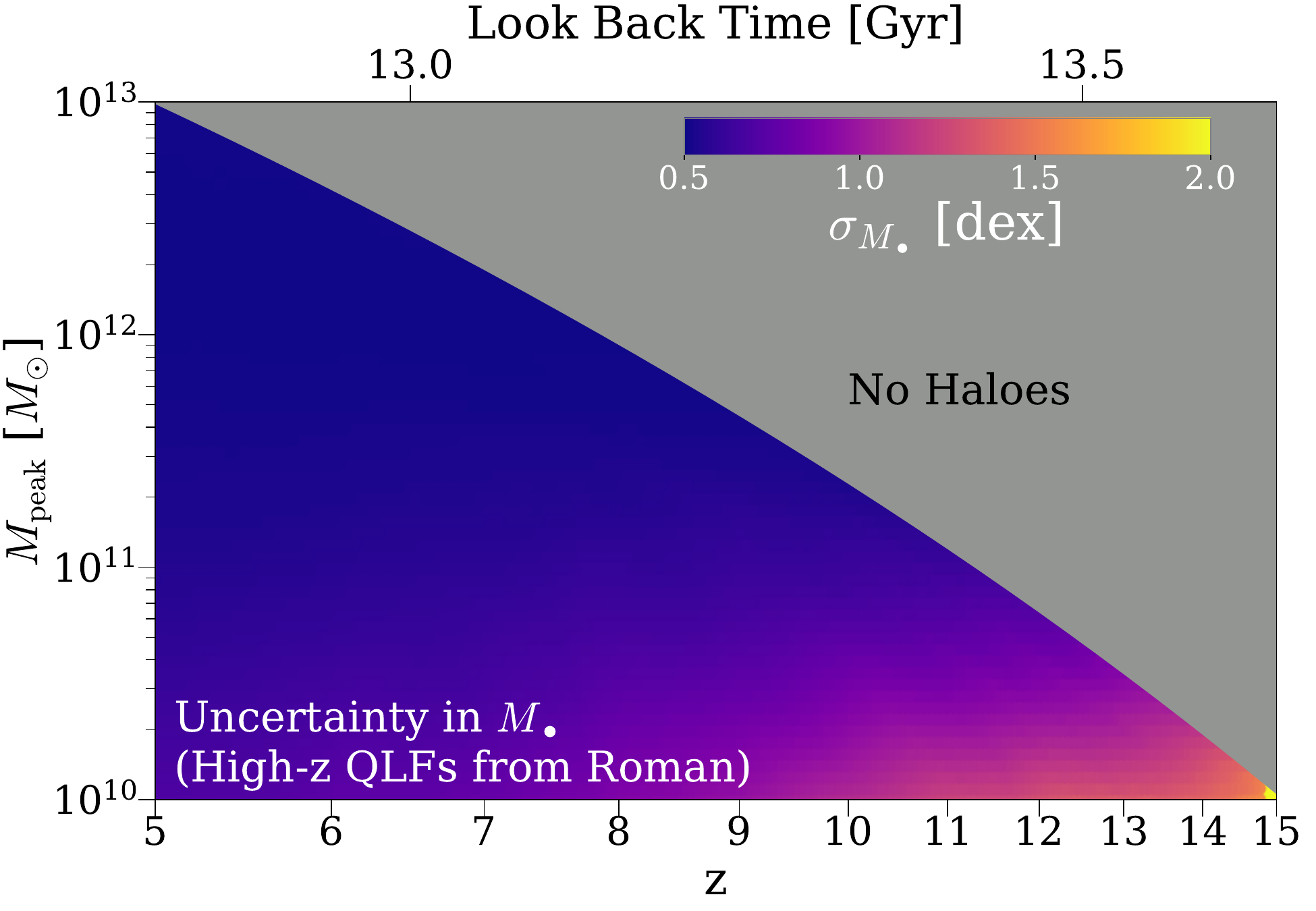}
}

\caption{\textbf{Top Panel: } The uncertainty in black hole mass ($M_\bullet$), as a function of \mpeak{} and $z$, with no quasar luminosity function (QLF) constraints between $z=6-10$. \textbf{Bottom Panel:} same as the top panel, but with $z=6-10$ mock QLF constraints from Roman High Latitude Wide Area Survey ($\sim 2000$ deg$^2$). The uncertainties are the quadratic sums of the those from MCMC process and the typical \mbh{} uncertainty from virial estimates, which is 0.5 dex.}
\label{f:mbh_uncertainty}
\end{figure}

In the top panel of Fig.\ \ref{f:mbh_uncertainty}, we show the uncertainty in \mbh{} as a function of \mpeak{} and $z$ from the fiducial \textsc{Trinity} model, where no SMBH data at $z\gtrsim 6$ were included to constrain our model. The \mbh{} uncertainty is a quadratic sum of: 1) the statistical uncertainty from the MCMC algorithm, which reflects the constraining power of all the data points when their values and uncertainties are perfectly accurate; and 2) the typical systematic uncertainty in \mbh{} mass measurement from virial estimates, i.e., 0.5 dex. Due to the lack of high-redshift SMBH data, the \mbh{} uncertainty increases significantly towards higher redshifts at $z\gtrsim 6$, reaching $\sim 1$ dex at $z\sim 10$ and $\gtrsim 2$ dex at $z\gtrsim12$.

We then added mock $z=6-10$ QLFs to see whether/how much these data can help constrain the halo--galaxy--SMBH connection at $z\gtrsim 6$. These QLFs are extrapolated based on the modeling (global fit B) of $z\lesssim 6$ QLFs by \citet{Shen2020}. These extrapolated QLFs are based on obscuration-corrected lower-redshift QLFs, so in principle include both obscured and unobscured AGNs. We determine the upper and lower luminosity limits of the QLFs assuming the characteristics of Roman High Latitude Wide Area Survey (HLS, $\sim2000$ deg$^2$); and 2) the Euclid Wide Survey (EWS, $\sim14500$ deg$^2$). To calculate the minimum luminosities of the mock QLFs, we convolved the quasar SED from \citet{Temple2021} with the filter transmission curves, and compare the fluxes with the near infrared magnitude limits of HLS (26.5 mag, \citealt{Wang2022}) and EWS (24.5 mag, \citealt{Euclid2022}). This yields minimum bolometric quasar luminosities of $10^{44.7}$ and $10^{45.5}$ erg/s for HLS and EWS, respectively. We truncate the bright end of mock QLFs at the luminosities where only one quasar will be detected in the redshift-luminosity bins. This maximum luminosity value is further capped at $10^{48}$ erg/s, which is the Eddington luminosity for a $10^{10} M_\odot$ SMBH. For HLS (EWS), the maximum luminosity is $10^{48}$ ($10^{48}$) erg/s at $z=6$, and $10^{46.5}$ ($10^{47.5}$) erg/s at $z=10$. 

As shown by the bottom panel of Fig.\ \ref{f:mbh_uncertainty}, the $z=6-10$ QLFs from HLS will greatly reduce the \mbh{} uncertainty in the early Universe. In particular, the total \mbh{} uncertainties are now dominated by the virial estimate uncertainties in \mbh{} for haloes with $M_\mathrm{peak}\gtrsim 10^{11} M_\odot$ at $z\lesssim10$. This means that $z=6-10$ QLFs strongly constrain SMBH growth history between $z=6-10$, when they are combined with (the high-redshift extrapolation of) all the other existing data included by \textsc{Trinity}. $z=6-10$ SMBH growth histories are so strongly constrained that we can precisely calculate $z=6-10$ SMBH masses by rewinding the growth histories of their $z\lesssim6$ descendants. As such, our knowledge about $z=6-10$ SMBH masses only depends on the accuracy of their $z\lesssim 6$ descendant SMBH masses. We do not show the results with EWS, because there is only very slight improvement from HLS to EWS. This improvement comes from the better sky coverage of Euclid, which provides better statistics for bright quasars.

\begin{figure}
\subfigure{
\includegraphics[width=0.48\textwidth]{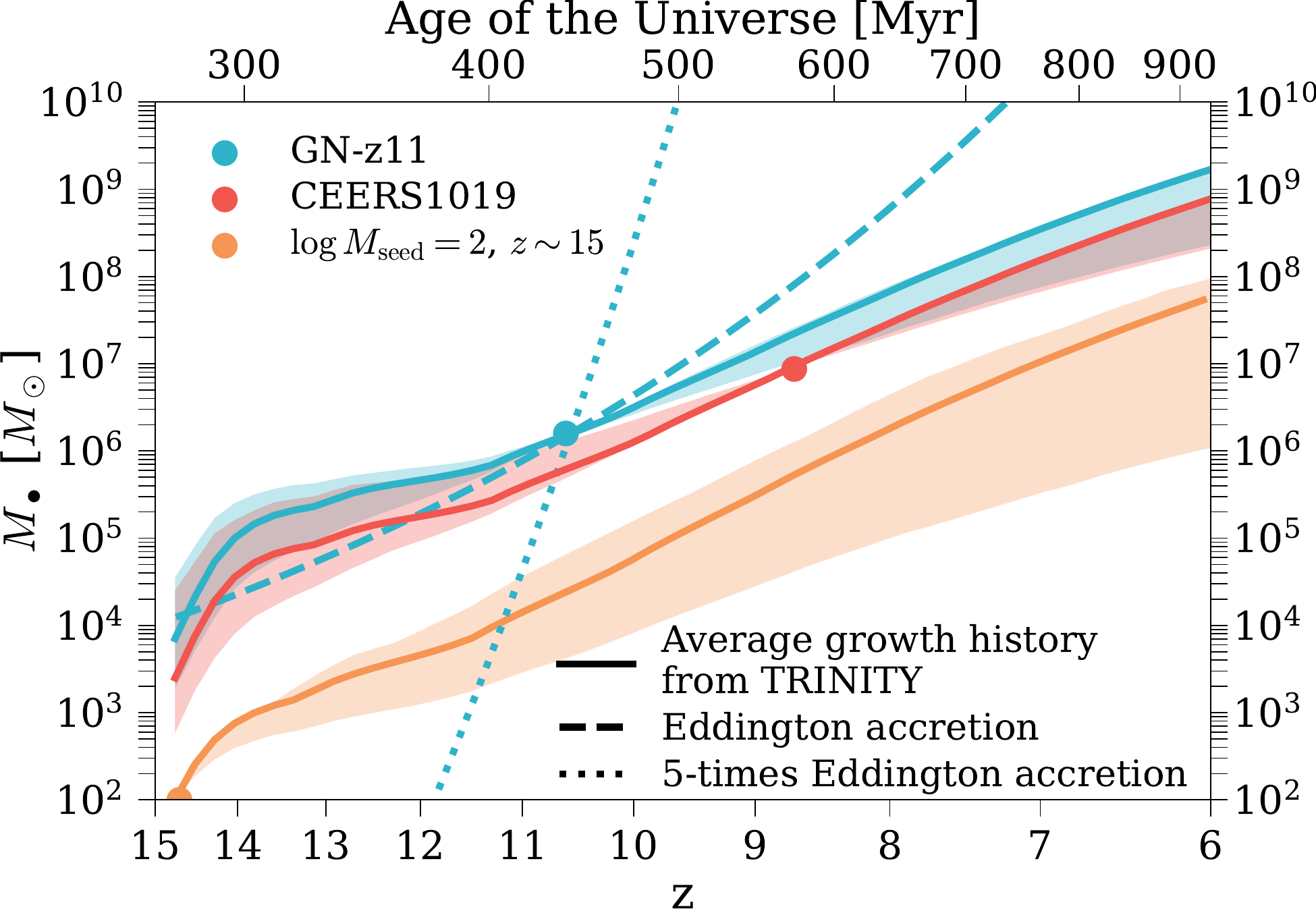}
}
\caption{Similar \emph{average} SMBH mass growth curves to Fig.\ \ref{f:growth_track_larson_maiolino}, but with extra constraints from simulated $z=6-10$ QLFs from Roman High Latitude Wide Area Survey. See \S\ref{ss:discussion_high_z_constraints}.}
\label{f:growth_track_larson_maiolino_high-z}
\end{figure}

The inclusion of high-redshift QLFs also does not significantly improve the constraints on early SMBH growth histories. In Fig.\ \ref{f:growth_track_larson_maiolino_high-z}, we show the average SMBH mass growth curves for the same objects as in Fig.\ \ref{f:growth_track_larson_maiolino}, but now with simulated $z=6-10$ QLFs as additional constraints. The uncertainties in both progenitor and descendant masses of these objects only decrease slightly. This is because the redshift dependence of AGN Eddington ratio distributions is already well constrained (see \S\ref{ss:justification}), and the addition of QLFs ends up mostly constraining SMBH mass. Nonetheless, we do caution that the high-redshift AGN Eddington ratios in \textsc{Trinity} are effectively extrapolated from low-redshift data. The shaded regions shown in Figs.\ \ref{f:growth_track_larson_maiolino} and \ref{f:growth_track_larson_maiolino_high-z} thus only represent the statistical uncertainties from the MCMC process, but do not include the systematic uncertainties from our input assumptions for such extrapolations, e.g., the exact parameterization of Eddington ratio distribution shape as a function of redshift.

\newpage
\section{Conclusions}
\label{s:conclusions}

In this work, we updated the \textsc{Trinity} model of the halo--galaxy--SMBH connection, by adding the latest $9\lesssim z \lesssim 13$ galaxy UV luminosity functions from JWST to our galaxy+SMBH data compilation. This additional constraint does not qualitatively change results presented in previous \textsc{Trinity} papers \citep{Zhang2021,Zhang2023,Zhang2023c}, but does increase the number of galaxies and SMBHs at fixed masses beyond $z\sim 9$. With the updated \textsc{Trinity} model, we predict the $z>6$ \bhsm{} relation for AGNs, the number density of massive ($M_\bullet>10^{9}\Msun$) SMBHs, and the host halo mass of massive SMBHs at $z\sim 6$.  \textbf{Key results are as follows}:
\begin{itemize}
    \item From $7\lesssim z \lesssim 10$, the median \bhsm{} relation for AGNs ($\log L_\mathrm{bol} \geq 45$) is biased higher than the intrinsic relation for \emph{all} SMBHs, due to Lauer bias. The \bhsm{} relation for quasars is also roughly redshift-independent. $z\gtrsim 7$ AGNs detected by JWST are broadly consistent with our predicted \bhsm{} relation given their AGN luminosities and mass measurement uncertainties.  (\S\ref{ss:bhsm_bright_quasars}, Fig.\ \ref{f:bhsm_bias}).

    \item The cumulative numbers of very massive ($M_\bullet > 10^9,\ 10^{10} M_\odot$) black holes above a certain redshift over the whole sky increase by six orders of magnitude from $z\sim 10$ to $z\gtrsim 2$, due to fast SMBH growth. Below $z\sim 2$, the SMBHs numbers increase slowly, both because SMBH growth slows down and the comoving observable volume is less. Including observational scatter in measured \mbh{} inflates the numbers of very massive black holes by a factor of $\sim 3-40$ due to Eddington bias. This overestimation is stronger for more massive black holes at lower redshifts, where the SMBH mass functions are steeper. (\S\ref{ss:results_huge_bh}, Figs.\ \ref{f:huge_bh}-\ref{f:bhmf}). 

    \item  $z\sim 6$ bright quasars with $M_\bullet > 10^{9}/10^{10} M_\odot$ are likely living in haloes with $M_\mathrm{peak} \sim (2-3)/(3-5) \times 10^{12} M_\odot$, which are $\sim 4.5-5.5 \sigma$ peaks in Gaussian random fields. The $z=0$ relics of these quasars are typically hosted by haloes with $M_\mathrm{peak} \gtrsim 3 \times 10^{14} M_\odot$. However, these host haloes only make up $\lesssim 1\%$ of all the haloes at this mass scale, making it very difficult to find the relic SMBHs at $z\sim 0$. We expect to find only $\lesssim 1$ such relic SMBH within $z\lesssim 0.05$ (\S\ref{ss:desc_big_mbh_at_z6}, Fig.\ \ref{f:prog_desc_mbh9z6}).

\end{itemize}

We also compared our predictions at $z\gtrsim 9$ with the two highest-redshift AGNs detected by JWST, GN-z11 ($z=10.6$) and CEERS\_1019 ($z=8.7$). Our main findings are:

\begin{itemize}

    \item The SMBH masses and Eddington ratios of GN-z11 and CEERS\_1019 are consistent with \textsc{Trinity}'s prediction that Eddington to super-Eddington AGNs with $6 \lesssim \log M_\bullet \lesssim 7$ dominate the quasar luminosity function at $\log L_\mathrm{bol}\sim 45$ and $z\sim9-10$ (\S\ref{sss:z9_z11_mbh_eta}, Fig.\ \ref{f:qlf_mbh_eta_z10}).

    \item The host halo masses are estimated to be $\sim (2-3)\times 10^{11} M_\odot$ and $\sim (3-6)\times 10^{11} M_\odot$ for GN-z11 and CEERS\_1019, respectively. These halo masses correspond to $\gtrsim 7\sigma$ peaks in the density field at those redshifts, which are extremely biased regions of the Universe. The $z=0$ descendants of GN-z11 and CEERS\_1019 are likely living in $(1-3)\times 10^{14} M_\odot$ haloes, but they are also very rare, similar to the local relics of $z=6$ quasars (\S\ref{sss:host_halo_larson_maiolino}, Fig.\ \ref{f:host_halo_larson_maiolino}). 

    \item On average, SMBHs grow at above(below) the Eddington rate above(below) $z\sim 13$. The predicted average growth histories make GN-z11 a potential progenitor of $z\sim 6$ quasars with $\log M_\bullet \gtrsim 9$, which is less likely to be the case for CEERS\_1019. Runaway mergers in nuclear star clusters represent a viable SMBH seeding mechanism for GN-z11 and CEERS\_1019 (\S\ref{sss:growth_track_larson_maiolino}, Fig.\ \ref{f:growth_track_larson_maiolino}). 

    \item The fraction of galaxies hosting AGNs with bolometric luminosities above $10^{45}$ erg/s, $f_\mathrm{gal}$ is a strong function of galaxy mass. At $z\gtrsim 6$, \textsc{Trinity} predicts that $f_\mathrm{gal}$ increases from $\lesssim 10^{-6}$ at $M_*\sim 10^9 M_\odot$ to $\sim 100\%$ at $M_*\sim 3\times 10^{10} M_\odot$ (\S\ref{sss:highz_agn_fraction}, Fig.\ \ref{f:highz_agn_frac}).

    \item The cumulative $z=8-12$ quasar luminosity function calculated from CEERS\_1019 and GN-z11 is not a strong enough constraint on the current \textsc{Trinity} model to cause significant shifts in the galaxy--SMBH connection. To impose a strong enough constraint on \textsc{Trinity}, more detections of AGNs above $L_\mathrm{bol}=10^{45}$ erg/s and/or better measurements of host galaxy mass would be needed. Alternatively, non-detection of such AGNs over a larger area would be necessary to confirm the prediction of AGN number densities by \textsc{Trinity}.

\end{itemize}

Finally, we showed that future $6<z<10$ AGN data will provide significant constraints on the early halo--galaxy--SMBH mass connection. Specifically, $z=6-10$ quasar luminosity functions (QLFs) measured by future wide area surveys (e.g., the Roman High Latitude Wide Area Survey and the Euclid Wide Survey) will reduce SMBH mass uncertainty from $\sim 1.0$ dex to $\sim 0.5$ dex for haloes with $M_\mathrm{peak}\gtrsim 10^{11} M_\odot$ at $z\lesssim10$. (\S\ref{ss:discussion_high_z_constraints}, Fig.\ \ref{f:mbh_uncertainty}).

\section*{Data availability}
\label{s:data_availability}

The parallel implementation of \textsc{Trinity}, the compiled datasets, and the posterior distribution of model parameters are available \href{https://github.com/HaowenZhang/TRINITY}{\textbf{online}}.

\section*{Acknowledgements}
\label{s:acknowledgements}

We thank Gurtina Besla, Frank van den Bosch, Haley Bowden, Alison Coil, Ryan Endsley, Sandy Faber, Nickolay Gnedin, Richard Green, Jenny Greene, Kate Grier, Melanie Habouzit, Kevin Hainline, Andrew Hearin, Yun-Hsin Huang, Raphael Hviding, Stephanie Juneau, David Koo, Andrey Kravtsov, Daniel Lawther, Jianwei Lyu, Chung-Pei Ma, Vasileios Paschalidis, Joel Primack, Eliot Quataert, George Rieke, Marcia Rieke, Jan-Torge Schindler, Xuejian Shen, Yue Shen, Dongdong Shi, Rachel Somerville, Fengwu Sun, Wei-Leong Tee, Yoshihiro Ueda, Marianne Vestergaard, Ben Weiner, Christina Williams, Charity Woodrum, and Minghao Yue for very valuable discussions.

Support for this research came partially via program number HST-AR-15631.001-A, provided through a grant from the Space Telescope Science Institute under NASA contract NAS5-26555. PB was partially funded by a Packard Fellowship, Grant \#2019-69646. PB was also partially supported by a Giacconi Fellowship from the Space Telescope Science Institute.  Finally, PB was also partially supported through program number HST-HF2-51353.001-A, provided by NASA through a Hubble Fellowship grant from the Space Telescope Science Institute, under NASA contract NAS5-26555.

Data compilations from many studies used in this paper were made much more accurate and efficient by the online \textsc{WebPlotDigitizer} code.\footnote{\url{https://apps.automeris.io/wpd/}} This research has made extensive use of the arXiv and NASA's Astrophysics Data System.

This research used the Ocelote supercomputer of the University of Arizona. The allocation of computer time from the UA Research
Computing High Performance Computing (HPC) at the University
of Arizona is gratefully acknowledged. The Bolshoi-Planck simulation was performed by Anatoly Klypin within the Bolshoi project of the University of California High-Performance AstroComputing Center (UC-HiPACC; PI Joel Primack).

The authors gratefully acknowledge the Gauss Centre for Supercomputing e.V. (www.gauss-centre.eu) and the Partnership for Advanced Supercomputing in Europe (PRACE, www.prace-ri.eu) for funding the MultiDark simulation project by providing computing time on the GCS Supercomputer SuperMUC at Leibniz Supercomputing Centre (LRZ, www.lrz.de).

\appendix


\bsp	
\label{lastpage}

{\footnotesize
\bibliography{trinity}
}

\end{document}